\documentclass[final]{siamltex}
\usepackage{epsf}
\usepackage{amssymb}
\usepackage{amsbsy}
\usepackage{verbatim}
\usepackage[sort&compress]{natbib} 
\begin{document}
\title{A Stochastic Immersed Boundary Method for 
Fluid-Structure Dynamics at Microscopic Length Scales}

\author{Paul J. Atzberger
\thanks{University of California, 
Department of Mathematics , Santa Barbara, CA 93106; 
e-mail: atzberg@math.ucsb.edu; phone: 805-679-1330;
Work supported by NSF Grant DMS - 9983646 and DMS-0635535.}
\and
Peter R. Kramer
\thanks{Rensselaer Polytechnic Institute, 
Department of Mathematics , Troy, NY 12180;
e-mail: kramep@rpi.edu; phone 518-276-6896;
Work supported by NSF CAREER DMS - 0449717.}
\and
Charles S. Peskin
\thanks{New York University, 
Department of Mathematics , New York, NY 10012;
e-mail: peskin@cims.nyu.edu.}
}

\maketitle

\begin{abstract}
In this work it is shown how the immersed boundary 
method of~\citep{peskin2002} for modeling flexible
structures immersed in a fluid can be extended to include thermal 
fluctuations.  A stochastic numerical method is 
proposed which deals with stiffness in the system
of equations by handling systematically
the statistical contributions of the fastest dynamics of the fluid and 
immersed structures over long time steps.  
An important feature of the numerical method is that 
time steps can be taken in which the degrees of freedom 
of the fluid are completely underresolved, partially resolved, or 
fully resolved while retaining a good level of 
accuracy.  Error estimates in each of these regimes are 
given for the method.  A number of theoretical and numerical checks are
furthermore performed to assess its physical fidelity.
For a conservative force, the method is found 
to simulate particles with the correct Boltzmann equilibrium 
statistics.  It is shown in three dimensions that the diffusion
of immersed particles simulated with the method has the correct 
scaling in the physical parameters.  The method is also shown to
reproduce a well-known hydrodynamic effect of
a Brownian particle in which the velocity autocorrelation function
exhibits an algebraic ($\tau^{-3/2}$) decay for long
times~\citep{sorensen2005,yp:tdcfm,bja:dvaf,
hinch1975, dorfman1970,ernst1971,hauge1973,mazur1974,cohen1974}.  
A few preliminary results are presented for more complex systems 
which demonstrate some potential application areas of the method.
\end{abstract}


\pagestyle{myheadings}
\thispagestyle{plain}
\markboth{P. ATZBERGER, P. KRAMER}
{STOCHASTIC IMMERSED BOUNDARY METHOD}

\section{Introduction}

In modeling many biological systems it is important to take into
account the interaction of flexible structures with a fluid. 
The immersed boundary method of~\citep{peskin2002} 
has found wide use as an efficient numerical method for simulating 
such systems.  Some examples include the study of blood flow 
around heart valves~\citep{peskin1997}, 
wave propagation in the inner ear~\citep{givelberg2001}, and 
the generation of lift in insect flight~\citep{miller2004}.
With experimental advances in molecular and cellular biology 
has come an increasing interest in developing methods to model 
qualitatively and quantitatively microscopic biological 
processes at the cellular and subcellular 
level~\citep{cpf:ccb,db:mc,jh:mmpc}.  The immersed boundary method 
provides a promising framework for simulating such systems.

At the cellular level the fluid may consist of either the aqueous
environment outside of the cell or the cytoplasm within.  Some
important flexible structures in the cellular context include the 
outer cell membrane, intracellular vesicles, cytoskeletal fibers,
and molecular motor proteins.  These structures 
play an important role in cell motility or cell division among other
processes~\citep{thecell}.  

In such systems the relevant features can span a range of length 
scales from tens of microns or more for the outer cell membrane and 
cytoskeletal fibers to tens of nanometers for individual 
cytoskeletal monomers and motor proteins.  At
these length scales thermal fluctuations of the system 
become significant and in many cases appear crucial to achieve 
biological function.  Some examples include
force generation and progression of molecular motors along 
cytoskeletal fibers~\citep{atzberger2005a,
jh:mmpc,
wittman2001}, osmotic 
effects such as vesicle 
and gel swelling~\citep{wolgemuth2004,go:doff,
evilevitch2003,
lipowsky2004,
gov2004}, 
and polymerization 
effects involved in force generating processes in cell
motility~\citep{go:fgcp,
jat:pm,
grimm2003,
robinson2004}.

Modeling such complex cellular systems in molecular detail is 
infeasible with methods such as molecular dynamics as 
a consequence of the immense computational cost required to resolve 
the broad range of active length and time scales.  This
suggests that a coarse-grained numerical approach must be taken
which does not resolve all of the detailed physics but rather 
attempts to make approximations that yield effective equations 
to capture the most relevant features of the dynamical phenomenon
being studied.  Here we discuss how the framework of the 
immersed boundary method can be extended for use in such
modeling by including thermal fluctuations to capture 
dynamical phenomena at the cellular length scale.  

The theory of nonequilibrium statistical mechanics indicates that the 
influence of thermal fluctuations on a mechanical system can typically 
be represented through the addition of thermal forcing terms which 
decorrelate rapidly in time.  The forcing can then be represented
by appropriate ``white noise'' processes.  This 
generally involves a nontrivial structure of correlations 
between the state variables in such a way that there is an 
energy balance between the thermal forcing and dissipation 
of the system so that a corresponding fluctuation-dissipation 
theorem for the system is satisfied ~\citep{IBrk:sp2,IBler:mcsp}.  

Several computational fluid dynamical schemes have been extended 
toward the microscale through such an inclusion of thermal forces.  
The most widely used approach is known as Stokesian or Brownian 
dynamics~\citep{jfb:sd,as:asds,dle:bdhi,ts:mms}. In this approach
the structures are modeled as collections of rigid 
``elementary particles'' which interact through force 
laws derived by approximating the fluid dynamics by a
quasi-steady Stokes flow.  This latter approximation is 
strictly appropriate only when, among other assumptions, the fluid 
density is much less than the density of the structures~\citep{jmd:cbmms}, 
a condition better met in engineering applications 
(such as suspensions~\citep{gb:dsss1,jfb:sd,as:asds}) than in 
physiological settings~\citep{lb:fsmab}.  The result of the 
underlying approximations in the Stokesian/Brownian dynamics method is a 
rather strongly coupled system of stochastic equations for the motion of 
the elementary particles.  For a well-designed computation the cost
of a simulation can be rendered roughly proportional to 
$ N \log N $, where $ N $ is the number of elementary 
particles~\citep{as:asds}.   
In the presence of fast 
time scales arising from thermal fluctuations and 
possibly chemically activated processes, the 
impact of  the quasi-steady Stokes 
approximation and
the representation of the elementary particles as rigid 
(rather than flexible)  on the accuracy of the 
simulation is not yet clear~\citep{jnr:bpdts}.

Another approach for modeling fluids with immersed structures
is Dissipative
Particle Dynamics~\citep{jba:dpdec,pe:dpdec,pe:smdpd,pjh:smhpd,cam:dpdef,
ip:scdpd,rdg:dpdbg,wko:nadpd,pn:hwyie}.  The method is built 
phenomenologically in terms of ``fluid particles''  which 
represent a parcel of fluid along with its collection of 
immersed structures.  When thermal forces are included in
the method the fluid particles are simulated with a stochastic 
system of equations modeling their (soft) interactions.  This
method however does not readily extend to the microscopic 
domain since the immersed structures within a parcel are not 
resolved in detail.  Dissipative Particle Dynamics may however
be appropriate for somewhat larger scale simulations in which 
one is interested in the effects of a numerous collection of 
immersed polymers or other structures on the dynamics of a fluid 
flow. 

A different class of approaches which emphasize the role of the fluid 
dynamics while making other simplifications has also been proposed.
These include finite-element~\citep{ns:dnsbm} and 
lattice-Boltzmann~\citep{ajcl:nspsd1} methods, in which the computational 
fluid dynamics are extended to include thermal forces in the 
fluid equations following the framework of~\citep{IBldl:ctp9}.
The immersed boundary method which we shall discuss belongs to this broad 
class of methods~\citep{kramer2003}.  A theoretical approach with certain
similarities to the immersed boundary method with thermal fluctuations 
has been proposed in~\citep{ottinger1989}, but differs 
in how thermal forces are treated for the immersed structures.
A virtue of the immersed boundary method when compared to other methods 
is the straightforward physical manner in which it 
approximates the interaction of the fluid with the flexible structures.

A key feature of the immersed boundary method, distinguishing 
it from Stokesian/Brownian dynamics, is that the dynamics of the fluid
are represented in the 
immersed boundary equations so that subtle inertial effects of the fluid can
be incorporated into the thermally fluctuating dynamics.  For example, 
as demonstrated in Subsection \ref{section_V_decay},  the method captures 
the slow decay ($\tau^{-3/2}$) in the tail of the autocorrelation function 
of the velocity 
of an immersed particle.  Another advantage of tracking the fluid dynamics 
is the natural way in which the immersed boundary method can respect the 
topology of flexible structures so that, for example, polymers do not cross 
themselves or each other.  This feature gives the immersed boundary method 
the potential for efficient simulation of polymer links and knots.

A basic description of how thermal fluctuations can be incorporated 
within the immersed boundary method was 
presented in~\citep{kramer2003}, and theoretical analysis of the 
physical behavior of the method through an asymptotic stochastic mode 
reduction calculation was developed in~\citep{prk:smrib}.  The immersed 
boundary method was found in these theoretical works to produce
generally the correct physical behavior for the thermal fluctuations of 
immersed structures.
 
Here we present a derivation for the thermal fluctuations of 
the immersed boundary method in the context of the 
time dependent Stokes equations.  We then present a new numerical
method developed from a novel time discretization of the stochastic equations.
In addition, further theoretical analysis of the framework is performed
to investigate the physical behavior of the method and
comparisons are made between theory and numerical simulations.

For stochastic differential equations most traditional finite 
difference methods, such as Runge-Kutta, achieve a lower order
of accuracy than for deterministic ordinary differential 
equations as a consequence of the nondifferentiability of Brownian
motion and its order ${t}^{1/2}$ scaling in 
time~\citep{platen1992}.  When considering the full system of 
equations of the immersed boundary method, 
 these issues 
are further compounded by a wide range of 
length and time scales that arise in many problems.  

For small length scale systems in which the Reynolds number is small and 
the fluid flow is Stokesian to a good approximation, the time scales
associated with the fine-scale fluid modes can be considerably
faster than the time scales of the large-scale fluid modes and the
immersed structures.
In many problems it is the dynamics of the immersed structures
and the large-scale features of the fluid flow that 
are of interest.  The fine-scale features of the fluid 
are incorporated in simulations primarily to determine their effects 
on the larger scales but are often not of direct interest in and of 
themselves.

Since the full system is rather stiff as a result of the fast time 
scales of the fine-scale fluid modes, we would like to be able to take 
time steps which underresolve those scales of the fluid which are not of 
primary interest in the simulation.  However, we must take care in how 
this is done, because otherwise the effects of the underresolved fluid 
modes on the more interesting degrees of freedom can be misrepresented.  
For example, the simple use of a method such as
Euler-Marayama~\citep{platen1992} leads to poor accuracy for the 
stochastic dynamics when
long time steps are used.  In this case, the particle trajectories are
found to be overly diffuse in the sense that the wrong scaling is
obtained for the mean squared distance traveled.

After a presentation of the general framework of the immersed boundary method 
without spatiotemporal discretization in 
Section~\ref{section_continuum_equ}, a stochastic 
analysis is developed in Section \ref{section_numerical_method} 
to design a numerical method which maintains accuracy over
time steps which can be longer than the time scales of some or
all of the fluid modes.   The systematically derived numerical scheme 
presented here improves upon in several ways a correction-factor approach 
taken in ~\citep{kramer2003} to achieve long time steps.
The numerical method is constructed from a new time 
discretization of the immersed boundary equations, in which  
the equations are integrated analytically 
using standard techniques from stochastic calculus under well 
controlled approximations.
The numerical method 
allows for the statistical 
contributions of the fast stochastic dynamics of the fluid, which 
are not explicitly resolved over long time steps, to be accounted 
for in the dynamics of the immersed structures.  Moreover,
the correlations in the statistics between the fluctuations of the 
degrees of freedom of the system are  handled systematically,
allowing for consistent realizations of the velocity field 
of the fluid and immersed structures to be simulated.

In Section \ref{section_accuracy},
 error estimates are given
which indicate that the method attains a good 
level of accuracy (in a strong statistical sense ~\citep{platen1992})
whether the fluid 
modes are completely underresolved, partially resolved, or fully 
resolved.  Only the degrees of freedom of the immersed structures constrain the time step.
The numerical method handles a
broad range of time steps in a unified manner, so that depending on the 
application,
the fast dynamics of the fluid can either be explicitly resolved or 
underresolved, with their effects correctly represented on the 
structural degrees of freedom.

In Section \ref{section_numerical_results},
an expression for the 
diffusion coefficient of immersed particles is derived.  The 
predictions are compared with the results of numerical simulations
showing good agreement with the theory for 
different particle sizes and both short and long time steps.  
It is further shown that 
for intermediate time steps, the method captures a well-known 
hydrodynamic effect of a Brownian particle, in which the decay
of the autocorrelation function of the velocity of the particle
decays algebraically ($\tau^{-3/2}$)
~\citep{sorensen2005,yp:tdcfm,bja:dvaf,
hinch1975, dorfman1970,ernst1971,hauge1973,mazur1974,cohen1974}.
Also in Section 
\ref{section_numerical_results}, numerical simulations are 
performed which confirm that the method produces the 
correct osmotic pressure and
equilibrium statistical distribution for the position of 
particles within an external potential at finite temperature.
To demonstrate more complex applications of the method, 
simulations are then presented which investigate 
behavior of the osmotic pressure associated with 
confinement of molecular dimers and polymer knots in a microscopic chamber, 
as well as   a basic model of a molecular motor 
protein immersed in a fluid subject to a hydrodynamic 
load force.

The physical consistencies we demonstrate 
in the method through our analysis and numerical 
experiments indicate that the stochastic immersed boundary 
method is a viable means to model on 
a coarse scale the influence of thermal fluctuations 
on the interaction of fluids and flexible structures.
This suggests that the method holds promise as an 
effective approach in modeling complex biological 
phenomena which operate at the cellular and 
subcellular level.  

\section{Fluid-Particle Equations}
\label{section_continuum_equ}

For the physical systems with which we shall be concerned,
the relevant length scales will typically be on the order of tens
of microns or smaller.   The amplitude of the velocity fluctuations on
these scales are sufficiently small, relative to the viscosity and length 
scale, 
that the Reynolds number is very small.   This will allow us to neglect 
the nonlinear advection term in the Navier-Stokes equation for 
the fluid dynamics.  However, we will not drop the time derivative term 
(as is often done in low Reynolds number limits~\citep{IBldl:ctp6}) 
because the dynamical time scales arising from Brownian motion and 
possibly certain vibrational modes of the immersed structures are in 
general too fast to allow this.  That is, the dynamics can generally 
exhibit time scales which are much shorter than the advection time 
scale (length scale divided by velocity scale), so we can drop the 
nonlinear advection term but not the time derivative of the velocity.
This leads us to the time-dependent Stokes equations for an incompressible 
fluid, which read
\begin{eqnarray}
\label{equ_cont_fluid_equ}
\rho{\frac{\partial\mathbf{u}(\mathbf{x},t)}{\partial{t}}} & = & 
\mu{\Delta}{\mathbf{u}(\mathbf{x},t)} 
-\nabla{p} + 
\mathbf{f}_{\mbox{\small total}}
(\mathbf{x},t) 
\\
\label{equ_cont_fluid_equ_incompressible}
\nabla\cdot{\mathbf{u}} & = & 0, 
\end{eqnarray}
where $p$ is the pressure arising from the incompressibility constraint,
$\rho$ is the fluid density, $\mu$ is the dynamic viscosity, and 
$ \mathbf{f}_{\mbox{\small total}} $ is the total force density 
acting on the fluid.  

The force
density acting on the fluid arises from two sources.  
The first source is the forces applied to the fluid by
the immersed structures and particles.  
This component of the force density,
denoted $\mathbf{f}_{\mbox{\small prt}}$, generally 
arises from the elastic deformations
of immersed structures, but they may also be applied
externally and transmitted by the immersed structures to
the fluid.
The second contribution to the force density is from the 
thermal fluctuations of the system and is denoted by 
$\mathbf{f}_{\mbox{\small thm}}$.  Each of these 
force densities will 
be discussed in greater detail below.  
Together, they comprise the total
force density acting on the fluid
\begin{eqnarray}
\mathbf{f}_{\mbox{\small total}}(\mathbf{x},t) 
& = & \mathbf{f}_{\mbox{\small prt}}(\mathbf{x},t) 
                     + \mathbf{f}_{\mbox{\small thm}}(\mathbf{x},t). 
\end{eqnarray}

The immersed boundary model for fluid-structure and fluid-particle 
coupling treats the flexible structures and particles to first
approximation as part of 
the fluid, representing their structural properties (such as
elasticity) through the  force density term 
$ \mathbf{f}_{\mbox{\small prt}}$
~\citep{peskin2002}.  
All structures, such as membranes,
polymers, and particles, are modeled as a collection of $M$ discrete 
``elementary particles,'' with locations denoted by $ \{\mathbf{X}^{[j]}
(t) \}_{j=1}^M $, 
which interact with force laws appropriate 
to their structural properties.  For simplicity, we consider 
the case in which all forces can be described in terms of 
a conservative potential $V(\{\mathbf{X}\})$ depending 
on the positions of the collection of elementary particles.  
More general force relations, including active forces, can be 
included~\citep{peskin2002}.  For notational convenience in the 
exposition,
the range of indices for the collection of elementary 
particles will often be omitted.  
We will often refer to the forces exerted by the immersed
structures as ``particle forces,'' since the structures are represented
in the numerical method as a collections of interacting particles. 

In the immersed boundary method, elementary
particles of size $a$ are represented by 
a function $\delta_a(\mathbf{x})$ which 
may be thought of as a Dirac delta function smoothed
over a length scale $a$ in such a manner that the smoothed
delta function has good numerical
properties (See Appendix~\ref{appendix_delta_func} and~\citep{peskin2002}).

This smoothed delta function is used
both in converting the force associated with an elementary particle to a
localized force density acting on the fluid:
\begin{eqnarray}
\mathbf{f}_{\mbox{\small prt}}(\mathbf{x},t)  =  
\sum_{j^{\prime} = 1}^{M} (-\nabla_{\mathbf{X}^{[j^{\prime}]}} 
V)(\{\mathbf{X}(t)\})\mbox{ }
                 \delta_{a}(\mathbf{x} - \mathbf{X}^{[j^{\prime}]}(t)) 
\end{eqnarray}
and in computing the velocity of the elementary particle by an
interpolation of the fluid velocity in its vicinity:
\begin{eqnarray}
\label{equ_X_dynamics}
\frac{d\mathbf{X}^{[j]}(t)}{dt} & = & \int_{\Lambda} 
              \delta_a(\mathbf{x} - \mathbf{X}^{[j]}(t))
          \mathbf{u}(\mathbf{x},t) d\mathbf{x}.
\end{eqnarray}
The integration is over the entire domain $\Lambda$ of the fluid.  
In the immersed boundary formulation, this includes the space
occupied by the immersed particles and structures, which are
thought of as parts of the fluid in which additional forces 
happen to be applied.  In particular, the domain $\Lambda$ is
independent of time, despite the motion of the immersed material.

We also note that in the present context the parameter $a$ is a
physical parameter of the model, since it is supposed to
represent a physical dimension of an elementary particle.
In particular, $ a $ is not a numerical parameter which is
supposed to vanish along with the meshwidth for the fluid 
computations as it is refined.
In this respect, the use of smoothed delta functions
described here is different from the standard use of such 
functions in immersed boundary computations.  The idea that 
smoothed delta functions could be used to model the physical
dimensions of immersed objects was previously proposed 
in \citep{meng1998} under the name "force cloud method".
%which has similarities to how our elementary 
%particles interact with the fluid.

\section{Numerical Method}
\label{section_numerical_method}
We shall now discuss a numerical discretization 
and specification of the thermal force density 
for the equations 
\ref{equ_cont_fluid_equ}--\ref{equ_X_dynamics} defining the immersed boundary method. 
A summary of the numerical method in algorithmic form
is given in Subsection \ref{section_summary_num_method}, 
followed by a heuristic discussion in 
Subsection \ref{section_heuristic_num_method}
and a mathematical derivation in 
Subsection \ref{section_method_derivation}.

\subsection{Summary of the Numerical Method}
\label{section_summary_num_method}
The numerical method is based upon a finite difference discretization of the differential equations 
\ref{equ_cont_fluid_equ}--\ref{equ_X_dynamics} describing the coupled dynamics of the fluid and the immersed structures.   The fluid variables (velocity field $ \mathbf{u} $ and pressure field $ p $) are represented on a periodic grid with length $L $ along each direction, $ N $ grid points along each direction, and grid spacing $ \Delta x = L/N $.  The values of these fields on the lattice will be denoted through subscripted variables such as $ \mathbf{u}_{\mathbf{m}} $ and $ p_{\mathbf{m}} $, where the  subscript $ \mathbf{m} = (m_1,m_2,m_3)$ is a vector with integer components indicating the grid point in question (relative to some  arbitrarily specified origin).    The position of the grid point with 
index $\mathbf{m}$ is denoted by $\mathbf{x}_{\mathbf{m}}$.  

The Discrete Fourier Transform (DFT) of the fluid variables plays an important role in the numerical simulation scheme, and is related to the physical space values on the grid through the formulas:\\\begin{eqnarray}
\hat{\mathbf{u}}_{\mathbf{k}} 
& = & \frac{1}{N^3} 
\sum_{\mathbf{m}} 
\mathbf{u}_{\mathbf{m}}
\exp\left({-i2\pi{\mathbf{k}}\cdot\mathbf{m}/N}\right) \\
\mathbf{u}_{\mathbf{m}} 
& = &
\sum_{\mathbf{k}} 
\hat{\mathbf{u}}_{\mathbf{k}}
\exp\left({i2\pi{\mathbf{k}}\cdot\mathbf{m}/N}\right),
\end{eqnarray}

where each of the sums in the above equations runs over 
the $N^3$ lattice
points defined by $0 \leq \mathbf{k}^{(\ell)}\leq N-1$,
and 
$0 \leq \mathbf{m}^{(\ell)} \leq N-1$, 
where the parenthesized superscripts $\ell=1,2,3$ denote the Cartesian components of the indicated vector.  In fact any translate of these blocks of lattice points could be used equivalently in the sums, due to the underlying periodicity. 

Time is discretized into time steps $ \Delta t $, and the values of the system variables at the $n^{\mathrm{th}}$ time step, corresponding to the time
$t_n = n \Delta{t}$, are denoted with a superscript integer $n$.  The procedure by which 
these variables are updated from one time step to the next is  now described:
%Each time step the following operations are performed to advance 
%the velocity field of the fluid and the configuration of the
%immersed structures, as represented by the positions of the elementary 
%particles.  In the notation $\Delta{t}$ denotes the time step and 
%$\Delta{x} = L/N$ denotes the grid spacing in each direction,
%where $L$ is the length of the periodic domain and $N$ is the 
%number of grid points in each direction.

\begin{enumerate}
\item The structural forces exerted by 
 the immersed structures are computed
and the lattice values of the 
particle force density field which is applied to the fluid is 
obtained from 
\begin{eqnarray}
\mathbf{f}^n_{\mathbf{m}}=  
\sum_{j^{\prime} = 1}^{N}-\left(\nabla_{\mathbf{X}^{[j^{\prime}]}} 
V\right)(\{\mathbf{X}^n\})\mbox{ }
                 \delta_{a}(\mathbf{x}_{\mathbf{m}} - \mathbf{X}^{n,[j^{\prime}]}).                 
\end{eqnarray}
Here and afterwards, we drop the subscript ``prt'' from the particle force density. 
The Fourier coefficients 
$\hat{\mathbf{f}}_{\mathbf{k}}^n$ of this particle force density field 
are computed 
using a Fast Fourier Transform (FFT).

\item The Fourier coefficients of the velocity field of the fluid are updated by the stochastic 
recurrence 
\begin{eqnarray}
\hat{\mathbf{u}}_{\mathbf{k}}^{n + 1}
& = & 
e^{-\alpha_{\mathbf{k}}\Delta{t}}\hat{\mathbf{u}}_{\mathbf{k}}^{n}
+ \frac{1}{\rho\alpha_{\mathbf{k}}}
\left(1 - e^{-\alpha_{\mathbf{k}}\Delta{t}}\right)\wp_{\mathbf{k}}^{\perp}
\hat{\mathbf{f}}_{\mathbf{k}}^n
+ \wp_{\mathbf{k}}^{\perp}
\hat{\boldsymbol{\Xi}}_{\mathbf{k}}^n,
\label{eq:velstochrec} 
\end{eqnarray} 
where $\wp_{\mathbf{k}}^{\perp}$ denotes the projection
orthogonal to $\mathbf{\hat{g}}_{\mathbf{k}}$ defined by
\begin{eqnarray}
\hat{\mathbf{g}}_k^{(j)} = 
\sin(2\pi\mathbf{k}^{(j)}/N)/{\Delta{x}}
\end{eqnarray}
which is used to enforce the incompressibility constraint
\ref{equ_cont_fluid_equ_incompressible}.
The factor 
$\hat{\boldsymbol{\Xi}}_{\mathbf{k}}^n = 
\sigma_{\mathbf{k}} \boldsymbol{\tilde{\boldsymbol{\eta}}}_{\mathbf{k}}$
accounts for the thermal fluctuations
over the time step, where  
$\boldsymbol{\tilde{\boldsymbol{\eta}}}_{\mathbf{k}}$ denotes 
a complex vector-valued random variable independent in 
$\mathbf{k}$,
 having independent real and imaginary components, 
each of which are Gaussian 
random variables with mean zero and variance one.
The variance  of $\hat{\boldsymbol{\Xi}}_{\mathbf{k}}^n$ is
determined in Subsection 
\ref{section_numerical_method_fluid} and is given by
\begin{eqnarray}
\sigma_{\mathbf{k}}^2 = 
\frac{D_{\mathbf{k}}}{\alpha_{\mathbf{k}}}
\left(1 - \exp\left(-2\alpha_{\mathbf{k}}\Delta{t}\right)\right), 
\label{eq:thermvar}
\end{eqnarray}
where 
%$\alpha_{\mathbf{k}}$ is defined in \ref{equ_DFT_def_alpha_k}
%and 
%$D_{\mathbf{k}}$ is defined in \ref{equ_thermal_forcing_strength}.
\begin{eqnarray}
\alpha_{\mathbf{k}} = \frac{2\mu}{\rho{\Delta{x}}^2}\sum_{j = 1}^{3} 
(1 - \cos(2\pi\mathbf{k}^{(j)}/N)))
\end{eqnarray}
and
\begin{eqnarray}
D_{\mathbf{k}} = 
\left\{
\begin{array}{ll}
\frac{k_B{T}}{\rho L^3} \alpha_{\mathbf{k}}  
& \mbox{, $\mathbf{k} \in \mathcal{K}$} \\
\frac{k_B{T}}{2\rho L^3} \alpha_{\mathbf{k}} 
& \mbox{, $\mathbf{k} \not\in \mathcal{K}$}
 \\
\end{array}
\right.
\end{eqnarray}
with
\begin{eqnarray}
\mathcal{K} = \left\{\mathbf{k}
%(\mathbf{k}^{(1)},\mathbf{k}^{(2)},\mathbf{k}^{(3)})
\mbox{ }| \mbox{ }\mathbf{k}^{(j)} = 0 \mbox{ }\mbox{ or } \mbox{ }
\mathbf{k}^{(j)} = N/2, \mbox{ }j = 1,2,3 \right\}.
\end{eqnarray}

\item The elementary particle
 positions are updated by
\begin{eqnarray}
\mathbf{X}^{{n+1},[j]} - \mathbf{X}^{n,[j]} & = & \sum_{\mathbf{m}} 
\delta_a(\mathbf{x}_{\mathbf{m}} - \mathbf{X}^{n,[j]})
\mathbf{\Gamma}_{\mathbf{m}}^n\Delta{x}^3, 
\end{eqnarray}
where $\mathbf{\Gamma}_{\mathbf{m}}^n$ is 
the time integrated velocity field of the fluid.
It is obtained by a discrete Inverse
Fast Fourier Transform (IFFT)
of appropriately generated random 
variables $\hat{\mathbf{\Gamma}}_{\mathbf{k}}^{n}$
in Fourier space:
\begin{eqnarray}
\mathbf{\Gamma}_{\mathbf{m}}^n 
= \int_{t_n}^{t_{n+1}} \mathbf{u}_{\mathbf{m}}(s) ds 
= \sum_{\mathbf{k}} \hat{\mathbf{\Gamma}}_{\mathbf{k}}^n
\cdot
\exp\left({{i2\pi{\mathbf{k}}\cdot \mathbf{m}}/{N}}\right).
\end{eqnarray}
The $\mathbf{\hat{\Gamma}}_{\mathbf{k}}^n$ are 
computed from
\begin{eqnarray}
\label{equ_gen_gamma_summary}
\mathbf{\hat{\Gamma}}_{\mathbf{k}}^n
& = & 
\mathbf{\hat{H}}_{\mathbf{k}} + 
c_{1,\mathbf{k}} \wp_{\mathbf{k}}^{\perp}
\mathbf{\hat{\boldsymbol{\Xi}}}^n_{\mathbf{k}}
+ c_{2,\mathbf{k}} \wp_{\mathbf{k}}^{\perp}
\mathbf{\hat{G}}_{\mathbf{k}}, 
\end{eqnarray}
where $\mathbf{\hat{\boldsymbol{\Xi}}}^n_{\mathbf{k}}$ is 
obtained from step 2 
 and 
$\mathbf{\hat{H}}_{\mathbf{k}}$ is computed from steps 
 1 and 2 by
\begin{eqnarray}
&& \\
\nonumber
\mathbf{\hat{H}}_{\mathbf{k}} & = & 
\frac{1 - 
\exp\left(-\alpha_{\mathbf{k}}\Delta{t}\right)}{\alpha_{\mathbf{k}}}
      \hat{\mathbf{u}}_{\mathbf{k}}^n
+ 
\left(
\frac{\Delta{t}}{\alpha_{\mathbf{k}}} 
+ 
\left(\frac{1}{\alpha_{\mathbf{k}}}\right)^2
\left(
\exp\left(-\alpha_{\mathbf{k}}\Delta{t}\right) - 1
\right)
\right)
\rho^{-1}
\wp_{\mathbf{k}}^{\perp}
\hat{\mathbf{f}}_{\mathbf{k}}^n. 
\end{eqnarray}
The random variable $\mathbf{\hat{G}}_{\mathbf{k}}$
is computed from scratch for each mode $\mathbf{k}$ 
by generating a complex vector-valued random variable 
having independent
real and imaginary components, each of which are Gaussian 
random variables 
with mean zero 
and variance one.
The constants in \ref{equ_gen_gamma_summary} are given by 
\begin{eqnarray}
c_{1,\mathbf{k}} & = & 
\frac{1}
{\alpha_{\mathbf{k}}}
\tanh\left(\frac{\alpha_{\mathbf{k}}\Delta{t}}{2}\right)
\end{eqnarray}
and
\begin{eqnarray}
c_{2,\mathbf{k}} & = & 
\sqrt{
\left(
\frac{2D_{\mathbf{k}}}{\alpha_{\mathbf{k}}^3}
\right)
\left(
\alpha_{\mathbf{k}}\Delta{t}
-2\tanh\left(\frac{\alpha_{\mathbf{k}}\Delta{t}}{2}\right)
\right).
} 
\end{eqnarray}

In this manner
the time integrated velocity field is consistently generated 
with the correct correlations with
$\{\mathbf{u}_{\mathbf{k}}^n\}$ and
$\{\mathbf{u}_{\mathbf{k}}^{n+1}\}$ from steps 
 1 and 2.  For more details, 
  see Subsections
\ref{section_numerical_method_particle}
and \ref{section_method_consistently_gen_modes}.
%This procedure is discussed in greater detail in Subsections 
%\ref{section_numerical_method_particle}
%and 
%\ref{section_method_consistently_gen_modes}.

\end{enumerate}
The computational complexity of the method, when
excluding the application specific forces acting
on the immersed structures, 
is dominated by the FFT and IFFT, 
which for a three 
dimensional lattice requires $O(N^3\log(N))$ arithmetic 
steps.

\subsection{Heuristic Discussion of the Numerical Method}
%changed a_k^{-1} to 1/a_k to be consistent with other sections
\label{section_heuristic_num_method}
We now briefly discuss each step of the numerical
scheme to give some intuition into how the 
method operates.  A more rigorous mathematical discussion and derivation is 
given in Subsection~\ref{section_method_derivation}.

The first step of the numerical scheme computes the 
structural forces exerted by the elementary particles as a function of their configuration,
and computes the discrete Fourier transform of the 
corresponding force density
field acting on the fluid.  
The second step updates
the fluid velocity in Fourier space by integrating over
the structural forces and thermal forces experienced over a time step.   The appearance of the time step $ \Delta t $ in exponential factors is due to the design of the method to maintain accuracy even if the fluid dynamics are partially resolved or underresolved by the time step, as in ``exponential time differencing'' schemes~\citep{eh:ltecn,bga:ltsmo,trefethen2005}.   The key time scale paired against the time step in these formulas is $1/\alpha_{\mathbf{k}} $, which describes the time scale of viscous damping of the Fourier mode $ \mathbf{k} $ of the fluid, as simulated by the numerical method.  
%It takes the form $ \alpha_{\mathbf{k}} \sim 4 \pi^{2} \mu^{2} |\mathbf{k}|^{2}/(\rho L^{2}) $ corresponding to a continuum limit when $ |\mathbf{k}| \ll N $ so that the Fourier mode in question describes variations on a length scale $ L/|\mathbf{k}| $ large relative to the grid spacing $ \Delta x = L/N $.  But for higher wavenumbers, the viscous dissipation rate of the modes is influenced by the finite difference approximation.
The first term of the stochastic recurrence equation~\ref{eq:velstochrec}
represents the viscously dissipated contribution of the fluid velocity from the previous time step.  The second term represents the contribution from the structural forces during the time step.  Since the elementary particle dynamics are assumed to be resolved by the time step, the structural force density itself appears in a simple way as effectively constant over the time step.  The response of the fluid velocity to this force has an exponential dependence on the time step to account for the possible levels of resolution of the viscous damping.  
The third term accounts for the thermal fluctuations over the time step through a mean zero Gaussian random variable 
$\hat{\boldsymbol{\Xi}}_{\mathbf{k}}^n$ with variance 
$ \sigma_{\mathbf{k}}^{2}$
describing the magnitude of the 
net contribution of the thermal fluctuations to the fluid velocity 
over the time step, see equation \ref{eq:thermvar}. % changed ref. to be consistent
Note that for time steps $\Delta t $ long compared to the relaxation time $1/\alpha_{\mathbf{k}}$ of the velocity Fourier mode, the variance approaches the constant value $ k_{B} T/(\rho L^{3}) $ corresponding to the equilibrium equipartition value.   On the other hand, for $ \Delta t \ll 1/\alpha_{\mathbf{k}}$, the variance of the thermal velocity increment is proportional to $ \Delta t $, so the magnitude of the increment is proportional to $ \sqrt{\Delta t} $.  This latter scaling is typical for the response of physical systems to noise driven by a large number of weak inputs (i.e., molecular fluctuations)~\citep{platen1992}.   
In both the second and third terms, the projection $\wp_{\mathbf{k}}^{\perp}$ enforces the 
incompressibility of the fluid.  The distinction in the definition of the factor $ D_{\mathbf{k}} $
with respect to wavenumbers $ \mathbf{k} $, as specified by the set $ \mathcal{K} $, is a purely 
technical issue related to the discrete Fourier transform; 
see Subsection~\ref{section_thermal_forcing}.

The third step of the numerical method updates 
the positions of the elementary particle positions composing the
immersed structures.  The random
variable $\boldsymbol{\Gamma}_{\mathbf{m}}^{n}$ 
represents the fluid velocity at lattice point $ \mathbf{x}_{\mathbf{m}}$ 
integrated over the time step, and is generated
in Fourier space using a procedure which also ensures that the immersed 
structures move with the correct correlations with the
previously computed fluid velocity values
$\mathbf{\hat{u}}_{\mathbf{k}}^{n}$ and 
$\mathbf{\hat{u}}_{\mathbf{k}}^{n+1}$.  The formulas defining 
$ \boldsymbol{\Gamma}_{\mathbf{m}}^{n} $ arise from an exact 
formula for integrating the fluid velocity over a time step, 
under the assumption that the structural forces can be treated 
as constant over the time step.

\subsection{Derivation of the Numerical Method}
\label{section_method_derivation}
The derivation first considers a spatial 
discretization of equation
\ref{equ_cont_fluid_equ} while leaving the 
system of equations continuous in time to avoid 
technical issues associated with the 
continuum formulation of the stochastic immersed 
boundary method with thermal forcing~\citep{prk:smrib}.
Since the equations are meant to serve as 
a physical model for the dynamics of the immersed structures
and fluid, a thermal forcing is derived for the semi-discretized 
system which is consistent with equilibrium statistical mechanics
in Subsection \ref{section_thermal_forcing}.
The time discretization of the numerical method for both the
dynamics of the fluid and immersed structures is then 
discussed in 
Subsection \ref{section_numerical_method_fluid}
and 
Subsection \ref{section_numerical_method_particle}.
The method takes special care to account for
correlations between the dynamics of the fluid and immersed structures,
which is discussed in Subsection 
\ref{section_method_consistently_gen_modes}.  
We remark that throughout the derivation, the
integration steps of the numerical method are designed
to maintain accuracy even when the time 
step does not fully resolve the dynamics 
of the fluid modes.

\subsubsection{Semi-discretization}

\label{derivation_semi_discr}
The equations \ref{equ_cont_fluid_equ} and 
\ref{equ_cont_fluid_equ_incompressible}
can be
discretized in space by finite difference approximations for the 
spatial derivatives
\citep{peskin2002}  
\begin{eqnarray}
\rho\frac{d\mathbf{u}_{\mathbf{m}}^{(\ell)}}{dt} 
& = & \mu
\sum_{q=1}^{3} 
\label{semidiscretization_fluid_equ1}
\frac{
\mathbf{u}_{\mathbf{m} - \mathbf{e}_{q}}^{(\ell)}(t) 
- 2\mathbf{u}_{\mathbf{m}}^{(\ell)}(t) 
+ \mathbf{u}_{\mathbf{m} + \mathbf{e}_{q}}^{(\ell)}(t)}
{\Delta{x}^2}  \\
& - &
\nonumber
%\sum_{\ell = 1}^{3}
\frac{p_{\mathbf{m} + \mathbf{e}_{\ell}}
- p_{\mathbf{m} - \mathbf{e}_{\ell}}}{2\Delta{x}}
%\mathbf{e}_{\ell}
+ \mathbf{f}_{\mbox{\small total}}^{(\ell)}(\mathbf{x}_{\mathbf{m}},t)  
\end{eqnarray}
\begin{eqnarray}
\sum_{\ell = 1}^{3}
\frac{\mathbf{u}_{\mathbf{m} + \mathbf{e}_{\ell}}^{(\ell)}(t) 
- \mathbf{u}_{\mathbf{m} - \mathbf{e}_{\ell}}^{(\ell)}(t)}{2\Delta{x}} & = & 0,
\end{eqnarray}
where $\mathbf{e}_{\ell}$ denotes the standard basis vector 
with all zero entries except for a one in the ${\ell}^{th}$ position. 
The parenthesized superscripts denote the vector
component.

The equations for the fluid-particle coupling become
\begin{eqnarray}
\label{semidiscretization_particle_equ0}
\mathbf{f}_{\mbox{\small prt}}(\mathbf{x}_{\mathbf{m}},t) & = & 
\sum_{j^{\prime} = 1}^{M}
-(\nabla_{\mathbf{X}^{[j^{\prime}]}} {V})(\{\mathbf{X}(t)\})
\mbox{ }
\delta_{a}
(\mathbf{x}_{\mathbf{m}} - \mathbf{X}^{[j^{\prime}]}(t)) \\
\label{semidiscretization_particle_equ1}
\frac{d\mathbf{X}^{[j]}(t)}{dt} & = & \mathbf{U}(\mathbf{X}^{[j]}(t),t) \\
\label{semidiscretization_avg_velocity_equ1}
\mathbf{U}(\mathbf{x},t) & = & \sum_{\mathbf{m}} 
             \delta_a(\mathbf{x}_{\mathbf{m}} - \mathbf{x})
             \mathbf{u}(\mathbf{x}_{\mathbf{m}},t)\Delta{x}^3.
\end{eqnarray}

Since we do not take the limit $a\rightarrow0$ 
as the meshwidth is refined (see above),
$\mathbf{U}(\mathbf{X}^{[j]}(t),t)$
does not become the same as $\mathbf{u}(\mathbf{X}^{[j]}(t),t)$,
even in the limit $\Delta{x} \rightarrow 0$.
That is, we are not simply evaluating the fluid
velocity at $\mathbf{X}^{[j]}(t)$, but instead averaging 
it over a region of width determined by the parameter $a$.  This averaging procedure ensures that the particle velocity remains finite and well-defined in the continuum limit, in which the pointwise values of the thermally fluctuating fluid velocity diverge.  Indeed, according to general statistical mechanical principles for continuum fields~\citep{IBldl:ctp9,IBrk:sp2,IBler:mcsp} (and not from any feature particular to the numerical method), the thermally fluctuating fluid velocity field manifests increasingly wilder fluctuations on smaller scales, and in the theoretical continuum limit approaches a sort of white noise structure.
%(actually blue noise) took out blue noise, might be a little confusing to reviewers, although I understand what you meant
This is not inherently problematic for physical interpretation, which only requires that meaningful values be obtained from averages over finite volumes, such as the size of a probe or an immersed structure.  Such averages are indeed finite, as can be understood intuitively through central limit theorem considerations by viewing them as averages of a large number of independent mean zero random variables due to the rapid spatial decorrelation of the noisy continuum velocity field.  From a more mathematical standpoint,  an application of the convolution theorem to the definition of $ \mathbf{U} $ in (\ref{semidiscretization_particle_equ0}) yields that $ \mathbf{U} $ can be represented as a nicely convergent Fourier series due to the decay of the Fourier coefficients of the smooth function $ \delta_{a} $.  

To obtain other desirable behaviors,
we also remark that it is important that we use the 
same weight function $\delta_a$ in averaging 
the fluid velocity as we do in applying force to the fluid, since this 
ensures that energy is properly conserved in the fluid-particle 
interaction.  With appropriate care in the construction of $\delta_a$, 
one can further ensure that momentum and angular momentum are conserved 
as well; see~\citep{peskin2002}.

\subsubsection{Fluid Equations in Fourier Space}
\label{section_fluid_fourier}

The Stokes equation is given in Fourier space by
\begin{eqnarray}
\label{equ_DFT_semi_discrete}
\frac{d\hat{\mathbf{u}}_{\mathbf{k}}}{dt} 
& = & 
-\alpha_{\mathbf{k}} \hat{\mathbf{u}}_{\mathbf{k}} 
- \mathrm{i} \rho^{-1} \hat{p}_{\mathbf{k}}\hat{\mathbf{g}}_{\mathbf{k}}
+ \rho^{-1} \hat{\mathbf{f}}_{\mbox{\small total},\mathbf{k}} \\
\label{equ_DFT_incompressible} 
\hat{\mathbf{g}}_{\mathbf{k}}\cdot\hat{\mathbf{u}}_{\mathbf{k}}& =& 0, 
\end{eqnarray}
where
\begin{eqnarray}
\label{equ_DFT_def_alpha_k}
\alpha_{\mathbf{k}} = \frac{2\mu}{\rho{\Delta{x}}^2}\sum_{j = 1}^{3} 
(1 - \cos(2\pi\mathbf{k}^{(j)}/N)))
\end{eqnarray}
\begin{eqnarray}
\label{equ_DFT_def_g_k}
\hat{\mathbf{g}}_k^{(j)} = 
\sin(2\pi\mathbf{k}^{(j)}/N)/{\Delta{x}}.
\end{eqnarray}

Since the velocity field of the fluid is real-valued, 
a further condition that must be satisfied by solutions 
of the equations 
\ref{equ_DFT_semi_discrete} - \ref{equ_DFT_incompressible}  is
\begin{eqnarray}
\label{equ_DFT_real_constr}
\overline{\hat{\mathbf{u}}_{\mathbf{N} - \mathbf{k}}} 
= \hat{\mathbf{u}}_{\mathbf{k}},
\end{eqnarray}
where $\mathbf{N}$ is shorthand for $(N,N,N)^T$ 
and the overbar denotes complex conjugation.   
Provided the force is real-valued, it can be shown 
that if this constraint holds for the initial conditions it will 
be satisfied for all time.

The Fourier coefficients $\hat{p}_{\mathbf{k}}(t)$ of the pressure 
need to be chosen in order to ensure that the incompressibility 
constraint is satisfied.  They can
be determined by projecting both sides of equation 
\ref{equ_DFT_semi_discrete} onto $\hat{\mathbf{g}}_{\mathbf{k}}$.  
By the incompressibility constraint 
\ref{equ_DFT_incompressible},
both of the terms involving $\hat{\mathbf{u}}_{\mathbf{k}}$ 
and its time derivative are zero under the projection.  
This gives at each time
\begin{eqnarray}
\label{equ_DFT_pressure}
\hat{p}_{\mathbf{k}}(t) & = & \frac{\mathrm{i}\hat{\mathbf{g}}_{\mathbf{k}}
\cdot\hat{\mathbf{f}}_{\mbox{\small total},\mathbf{k}}(t) }
{|\hat{\mathbf{g}}_{\mathbf{k}}|^{2}}. 
\end{eqnarray}
For those values of $\mathbf{k}$ that make 
$\hat{\mathbf{g}}_{\mathbf{k}} = 0$, the incompressibility constraint is 
trivial, and by convention we shall take $\hat{p}_{\mathbf{k}}(t) = 0$ 
for such $\mathbf{k}$.

For future reference,
let the projection in the direction 
$\hat{\mathbf{g}}_{\mathbf{k}}$ be denoted by
\begin{eqnarray}
\label{equ_def_proj_parallel}
\wp_{\mathbf{k}}^{\parallel} = 
\frac{\hat{\mathbf{g}}_{\mathbf{k}}\hat{\mathbf{g}}_{\mathbf{k}}^T}
{|\hat{\mathbf{g}}_{\mathbf{k}}|^2}
\end{eqnarray}
and the projection orthogonal to $\hat{\mathbf{g}}_{\mathbf{k}}$ be denoted by
\begin{eqnarray}
\label{equ_def_proj_perp}
\wp_{\mathbf{k}}^{\perp} = 
\left(
\mathcal{I} 
- \frac{\hat{\mathbf{g}}_{\mathbf{k}}\hat{\mathbf{g}}_{\mathbf{k}}^T}
{|\hat{\mathbf{g}}_{\mathbf{k}}|^2}.
\right)
\end{eqnarray}
For those modes for which $ \mathbf{\hat{g}}_{\mathbf{k}} = \mathbf{0} $, the 
corresponding projections will be understood to be defined 
$ \wp_{\mathbf{k}}^{\parallel} = \mathbf{0} $ 
and $ \wp_{\mathbf{k}}^{\perp} = \mathcal{I} $.  The set of 
indices on which $ \mathbf{\hat{g}}_{\mathbf{k}} = \mathbf{0} $ is 
given by
\begin{eqnarray}
\label{equ_def_K}
\mathcal{K} = \left\{\mathbf{k}
%(\mathbf{k}^{(1)},\mathbf{k}^{(2)},\mathbf{k}^{(3)})
\mbox{ }| \mbox{ }\mathbf{k}^{(j)} = 0 \mbox{ }\mbox{ or } \mbox{ }
\mathbf{k}^{(j)} = N/2, \mbox{ }j = 1,2,3 \right\}.
\end{eqnarray}

For a 
function $\mathbf{w}_{\mathbf{m}}$ defined over the 
discrete lattice sites indexed by $\mathbf{m}$, 
 the 
corresponding projection operations 
in physical space are given by
\begin{eqnarray}
(\wp^{\parallel} \mathbf{w})_{\mathbf{m}} & = & 
\sum_{\mathbf{k}} 
\wp_{\mathbf{k}}^{\parallel}
\hat{\mathbf{w}}_{\mathbf{k}}
\exp\left({i2\pi{\mathbf{k}}\cdot\mathbf{m}/N}\right)
\end{eqnarray}
and 
\begin{eqnarray}
(\wp^{\perp}\mathbf{w})_{\mathbf{m}} & = & 
\sum_{\mathbf{k}} 
\wp_{\mathbf{k}}^{\perp}\hat{\mathbf{w}}_{\mathbf{k}}
\exp\left({i2\pi{\mathbf{k}}\cdot\mathbf{m}/N}\right). 
\end{eqnarray}

\subsubsection{Thermal Forcing}
\label{section_thermal_forcing}
Following standard practice in nonequilibrium statistical
mechanics~\citep{IBrk:sp2,IBler:mcsp}, 
the thermal fluctuations of the system are modeled as Gaussian white
noise.  Formally, the Fourier coefficients of the thermal forcing 
can be written as 
\begin{eqnarray}
\hat{\mathbf{f}}_{\mbox{\small thm},\mathbf{k}} = \rho\sqrt{2D_{\mathbf{k}}}
\frac{d\mathbf{\tilde{B}}_{\mathbf{k}}(t)}{dt}. 
\end{eqnarray}
The factor $D_{\mathbf{k}}$ (which is to be specified) describes the strength 
of the thermal forcing of the 
$\mathbf{k}^{th}$ mode and $\mathbf{\tilde{B}}_{\mathbf{k}}(t)$ denotes
a complex-valued Brownian motion with the real and imaginary 
parts of each component consisting of an independent 
standard Brownian motion~\citep{platen1992}.  The dependence 
on $\mathbf{k}$ will be discussed below.

Standard Brownian motion $B(t)$ for our purposes will refer to 
the continuous stochastic process which is defined by the 
following properties:
\[
\begin{array}{ll}
\mbox{(i)}   & \mbox{$B(0) = 0$,
} \\
\mbox{(ii)} & \mbox{$E\left(B(t_2)- B(t_1)\right) = 0$,
} \\
\mbox{(iii)}  & 
\mbox{$E\left(\left|B(t_2)- B(t_1)\right|^2\right) = |t_2 - t_1|$, 
} \\
\mbox{(iv)}   & \mbox{The increments 
$B(t_2) - B(t_1)$ and $B(t_4) - B(t_3)$ are}  \\
      & \mbox{independent Gaussian random variables 
whenever $t_1<t_2\leq t_3<t_4$, 
}  \\
\end{array}
\]
where $E\left(\cdot\right)$ denotes the expected value.  
Standard Brownian motion in $d$ dimensions is defined
as a stochastic process where each vector component is an independent
one-dimensional Brownian motion.  For a further discussion of the 
properties of Brownian motion and related technical issues, see
\citep{oksendal2000} or~\citep{cwg:hsm}.

The discretized Stokes equation \ref{equ_DFT_semi_discrete} 
with only thermal forcing (no immersed structural 
 forces)
can be expressed in stochastic differential notation as \\
\begin{eqnarray}
\label{equ_stokes_DFT_thermal1}
d\hat{\mathbf{u}}_{\mathbf{k}} 
& = & 
\left[
-\alpha_{\mathbf{k}} \hat{\mathbf{u}}_{\mathbf{k}} 
-i\rho^{-1} \hat{p}_{\mathbf{k}}\hat{\mathbf{g}}_{\mathbf{k}} \right] dt 
+ \sqrt{2D_{\mathbf{k}}}d\mathbf{\tilde{B}}_{\mathbf{k}}(t) \\
\label{equ_stokes_DFT_thermal1_incomp}
\hat{\mathbf{g}}_{\mathbf{k}}\cdot \hat{\mathbf{u}}_{\mathbf{k}} & = & 0 \\
\label{equ_stokes_DFT_thermal1_real_contr}
\overline{\hat{\mathbf{u}}_{\mathbf{N} - \mathbf{k}}} 
& = & \hat{\mathbf{u}}_{\mathbf{k}}, 
\end{eqnarray}
where $d\mathbf{\tilde{B}}_{\mathbf{k}}(t)$ denotes increments of the
complex-valued Brownian motion associated with the $\mathbf{k}^{th}$ mode.
To ensure that the thermal forcing be real-valued, 
 the
Brownian increments are correlated in $\mathbf{k}$ by 
the constraint 
\begin{eqnarray}
\label{equ_stokes_dB_contr}
\overline{d\mathbf{\tilde{B}}_{\mathbf{N} - \mathbf{k}}} 
 =  d\mathbf{\tilde{B}}_{\mathbf{k}}. 
\end{eqnarray}

As discussed in Section \ref{section_fluid_fourier}, the 
pressure can be expressed in terms of the force 
acting on the fluid using \ref{equ_DFT_pressure}.  
By formal substitution into 
\ref{equ_stokes_DFT_thermal1},
the incompressibility constraint can be 
incorporated through an appropriate projection 
operation which allows for the two equations 
\ref{equ_stokes_DFT_thermal1}
and 
\ref{equ_stokes_DFT_thermal1_incomp}
to be expressed as the single equation
\begin{eqnarray}
\label{equ_stokes_DFT_thermal2}
d\hat{\mathbf{u}}_{\mathbf{k}} 
+ \alpha_{\mathbf{k}} \hat{\mathbf{u}}_{\mathbf{k}}dt 
& = & 
\sqrt{2D_{\mathbf{k}}}
\wp_{\mathbf{k}}^{\perp}d\mathbf{\tilde{B}}_{\mathbf{k}}(t). 
\end{eqnarray}

Since the incompressibility constraint is equivalent to
$\wp^{\perp}_{\mathbf{k}} 
\hat{\mathbf{u}}_{\mathbf{k}} = \hat{\mathbf{u}}_{\mathbf{k}}$,
the constraint will be satisfied for all time provided it
holds at the initial time.  Consequently, when
$\hat{\mathbf{g}}_{\mathbf{k}} \not= \mathbf{0}$ 
($\mathbf{k} \not \in \mathcal{K}$ as defined in 
\ref{equ_def_K}),
the real and imaginary part of the stochastic process 
$\hat{\mathbf{u}}_{\mathbf{k}}(t)$ 
remain in the plane orthogonal to
$\hat{\mathbf{g}}_{\mathbf{k}}$ for all time.
When $\hat{\mathbf{g}}_{\mathbf{k}} = \mathbf{0}$
($\mathbf{k} \in \mathcal{K}$), 
no constraint
is imposed on the real part, but
the real-valuedness condition \ref{equ_DFT_real_constr} 
requires that the imaginary component vanish.

Equation \ref{equ_stokes_DFT_thermal2} can be solved by
the method of integrating factors to obtain
\begin{eqnarray}
\label{equ_int_factor_sol}
\hat{\mathbf{u}}_{\mathbf{k}}(t) 
& = & 
\sqrt{2D_{\mathbf{k}}}
\wp_{\mathbf{k}}^{\perp}
\int_{-\infty}^{t} 
e^{-\alpha_{\mathbf{k}}(t - s)}
d\mathbf{\tilde{B}}_{\mathbf{k}}(s). 
\end{eqnarray}
Since $\hat{\mathbf{u}}_{\mathbf{k}}(t)$ is the projection of 
an Ito integral with deterministic integrand, it is at each 
time $ t $ a  Gaussian random variable with mean zero. 

The variance of $\hat{\mathbf{u}}_{\mathbf{k}}(t)$ can be 
computed from \ref{equ_int_factor_sol} by
\begin{eqnarray}
\label{equ_exp_u_k_sq}
E\left(|\mathbf{\hat{u}}_{\mathbf{k}}|^2\right)
& = & 
\,\mbox{Tr}\, 
\left(E\left(
\mathbf{\hat{u}}_{\mathbf{k}}
\overline{\mathbf{\hat{u}}_{\mathbf{k}}}^T\right)
\right) \\
& = & 
\nonumber
2 D_{\mathbf{k}}
\,\mbox{Tr}\, 
\left(
\wp_{\mathbf{k}}^{\perp}
\int_{-\infty}^{t} 
\int_{-\infty}^{t} 
e^{-\alpha_{\mathbf{k}}(2t - s  - s')}
E\left(
d\mathbf{\tilde{B}}_{\mathbf{k}}(s)
\overline{
d\mathbf{\tilde{B}}_{\mathbf{k}}^T(s')} 
\right)
\right).  
\end{eqnarray}
To proceed further, we must distinguish between 
the cases 
$\mathbf{k} \in \mathcal{K}$ and 
$\mathbf{k} \not\in \mathcal{K}$,
where $\mathcal{K}$ is defined in \ref{equ_def_K}.

For $\mathbf{k} \in \mathcal{K}$, 
we have 
$\wp_{\mathbf{k}}^{\perp} = \mathcal{I}$
and from constraint \ref{equ_stokes_dB_contr} that
the Brownian motion $ \mathbf{\tilde{B}}_{\mathbf{k}} (s) $ 
is real valued.
Therefore,
\begin{eqnarray}
E\left(
d\mathbf{\tilde{B}}_{\mathbf{k}}(s)
\overline{
d\mathbf{\tilde{B}}_{\mathbf{k}}^T(s')} 
\right)
= \mathcal{I}\delta(s - s')ds ds', 
\end{eqnarray}
and it follows that 
\begin{eqnarray}
\label{equ_exp_u_k_sq_in_K}
E\left(|\mathbf{\hat{u}}_{\mathbf{k}}|^2\right)
= \frac{3 D_{\mathbf{k}}}{\alpha_{\mathbf{k}}}
\end{eqnarray}
when $\mathbf{k} \in \mathcal{K}$.  

For $\mathbf{k} \not\in \mathcal{K}$, the Brownian motion
is complex valued and $\wp_{\mathbf{k}}^{\perp}$ is 
a projection onto the two-dimensional subspace orthogonal
to $\mathbf{\hat{g}}_{\mathbf{k}}$.  Therefore, 
\begin{eqnarray}
E\left(
d\mathbf{\tilde{B}}_{\mathbf{k}}(s)
\overline{
d\mathbf{\tilde{B}}_{\mathbf{k}}^T(s')} 
\right)
= 2 \wp_{\mathbf{k}}^{\perp} 
 \delta(s - s') ds ds'
\end{eqnarray}
and 
\begin{eqnarray}
\,\mbox{Tr}\,
\left(
\wp_{\mathbf{k}}^{\perp}
\right)
= 2, 
\end{eqnarray}
from which it follows that
\begin{eqnarray}
\label{equ_exp_u_k_sq_not_in_K}
E\left(|\mathbf{\hat{u}}_{\mathbf{k}}|^2\right)
= \frac{4 D_{\mathbf{k}}}{\alpha_{\mathbf{k}}}
\end{eqnarray}
for $\mathbf{k} \not\in \mathcal{K}$.
We remark that the formal
calculations above can be justified rigorously by 
applying Ito's Isometry directly to equation
\ref{equ_exp_u_k_sq}; see reference \citep{oksendal2000}.

To determine $D_{\mathbf{k}}$, we shall now compare these 
results with those that are obtained if we impose the 
condition 
that
the immersed boundary method exhibit fluctuations governed by the
Boltzmann distribution, as required by classical statistical 
mechanics. 
By Parseval's Lemma,
 the total kinetic energy can
be expressed in terms of the Fourier modes of the fluid by
\begin{eqnarray}
\mathcal{E}[\{\mathbf{u}_{\mathbf{k}}\}] & = & 
                      \frac{\rho}{2} \sum_{\mathbf{m}} 
                      |\mathbf{u}_{\mathbf{m}}|^2 
                      \Delta{x}^3 \\
\nonumber
                 & = & \frac{\rho}{2} \sum_{\mathbf{k}} 
                      |\hat{\mathbf{u}}_{\mathbf{k}}|^2 L^3.
\end{eqnarray}
The density of the Boltzmann distribution is then given by
\begin{eqnarray}
\label{equ_density_boltz}
\tilde{\Psi}(\{\hat{\mathbf{u}}_{\mathbf{k}}\}) 
= \frac{1}{\tilde{Z}} \exp\left(-
\frac{
{\rho L^3}
\sum_{\mathbf{k}} 
|\hat{\mathbf{u}}_{\mathbf{k}}|^2}
{2k_B{T}
}\right), 
\end{eqnarray}
where $\tilde{Z}$ is the partition function obtained by integrating
$\hat{\mathbf{u}}_{\mathbf{k}}$ over the constrained subspace 
\begin{eqnarray}
\Omega = \{\{\hat{\mathbf{u}}_{\mathbf{k}}\} \mbox{  }| \mbox{  } 
          \hat{\mathbf{g}}_{\mathbf{k}} \cdot 
          \hat{\mathbf{u}}_{\mathbf{k}} = 0,
          \overline{\hat{\mathbf{u}}_{\mathbf{N} - \mathbf{k}}} 
           = \hat{\mathbf{u}}_{\mathbf{k}}
         \}.
\end{eqnarray}
Each degree of freedom of the fluid contributes a quadratic term 
to the energy of the system, 
 giving a Boltzmann distribution which
is Gaussian.  Therefore, the equipartition theorem holds 
and each independent degree of freedom contributes on average 
$\frac{1}{2} k_B {T}$ 
 to the kinetic energy. 

For a particular wavenumber $\mathbf{k} \in \mathcal{K}$, 
the mean contribution to the energy 
is 
\begin{eqnarray}
\frac{\rho L^3}{2}\frac{3 D_{\mathbf{k}}}{\alpha_{\mathbf{k}}}, 
\end{eqnarray}
where the expression \ref{equ_exp_u_k_sq_in_K} 
for $E\left(|\mathbf{\hat{u}}_{\mathbf{k}}|^2\right)$ has been used.
For such wavenumbers there are 3 independent degrees of freedom 
corresponding to
the 3 real components of $\mathbf{\hat{u}}_{\mathbf{k}}$.  By the 
equipartition theorem this requires 
\begin{eqnarray}
\frac{\rho L^3}{2}\frac{3 D_{\mathbf{k}}}{\alpha_{\mathbf{k}}} 
 = \frac{3}{2} k_B
 {T}, 
\end{eqnarray}
which gives 
\begin{eqnarray}
D_{\mathbf{k}} = \alpha_{\mathbf{k}} 
\frac{k_B 
T}{\rho L^3}
\end{eqnarray}
when $\mathbf{k} \in \mathcal{K}$.

For $\mathbf{k} \not\in \mathcal{K}$, we must consider
the pair $(\mathbf{k},\mathbf{N} - \mathbf{k})$ together,
since $\mathbf{\hat{u}}_{\mathbf{k}} = 
\overline{\mathbf{\hat{u}}_{\mathbf{N} - \mathbf{k}}}$.
The 
contribution to the mean energy of these two 
wavenumbers is 
\begin{eqnarray}
2\frac{\rho L^3}{2}\frac{4 D_{\mathbf{k}}}{\alpha_{\mathbf{k}}}, 
\end{eqnarray}
where the expression \ref{equ_exp_u_k_sq_not_in_K} 
for $E\left(|\mathbf{\hat{u}}_{\mathbf{k}}|^2\right)$ 
and $E\left(|\mathbf{\hat{u}}_{\mathbf{N} - \mathbf{k}}|^2\right)$ 
has been used.  The number of independent degrees of freedom corresponding 
to the pair of wavenumbers $(\mathbf{k},\mathbf{N} - \mathbf{k})$
is $4$, since the real vector space orthogonal to 
$\mathbf{\hat{g}}_{\mathbf{k}}$ is two-dimensional and 
$\mathbf{\hat{u}}_{\mathbf{k}}$ is complex valued.

The equipartition theorem in this case requires that 
\begin{eqnarray}
2\frac{\rho L^3}{2}\frac{4 D_{\mathbf{k}}}{\alpha_{\mathbf{k}}}
& = & \frac{4}{2} k_B 
{T}, 
\end{eqnarray}
which gives 
\begin{eqnarray}
D_{\mathbf{k}} = \alpha_{\mathbf{k}} 
\frac{k_B 
T}{2 \rho L^3}
\end{eqnarray}
when $\mathbf{k} \not\in \mathcal{K}$.

To summarize, the following fluctuation-dissipation 
relation \citep{IBrk:sp2,IBler:mcsp} is obtained
when considering the constraints 
\ref{equ_DFT_real_constr} and
\ref{equ_DFT_incompressible} 
imposed on the velocity field of the fluid: 
\begin{eqnarray}
\label{equ_thermal_forcing_strength}
D_{\mathbf{k}} = 
\left\{
\begin{array}{ll}
\frac{k_B{T}}{\rho L^3} \alpha_{\mathbf{k}}  
& \mbox{, $\mathbf{k} \in \mathcal{K}$} \\
\frac{k_B{T}}{2\rho L^3} \alpha_{\mathbf{k}} 
& \mbox{, $\mathbf{k} \not\in \mathcal{K}$}. 
 \\
\end{array}
\right.
\end{eqnarray}
% ******************************************************

We remark that the Fourier mode of the fluid associated with
$\mathbf{k} = [0,0,0]^T $ corresponds to translation of the 
fluid as a whole.  From \ref{equ_DFT_def_alpha_k}
the zero mode has $\alpha_{\mathbf{0}} = 0$, 
which indicates that the fluid has
no translational damping.  As a consequence of 
\ref{equ_thermal_forcing_strength} the mode 
$\mathbf{\hat{u}}_{\mathbf{0}}$ is not thermally forced, which can 
also be understood physically by the conservation of total momentum 
by the internal thermal fluctuations.  
Thus for a fluid initially at 
rest with no net external force on the fluid
as a whole, the translational mode remains zero 
$\mathbf{\hat{u}}_{\mathbf{0}} = \mathbf{0}$ under the thermal forcing.

\subsubsection{Numerical Method for the Fluid}
\label{section_numerical_method_fluid}
To deal with the significant range in 
time scales for the modes of the fluid and immersed structures,
we develop a time-stepping scheme that freezes the positions and 
forces exerted by the elementary particles over a time step $\Delta t$, but 
otherwise integrates the dynamical equations exactly.
With this
approximation the set of equations 
\ref{equ_stokes_DFT_thermal2} can be solved analytically 
using the methods of stochastic calculus 
\citep{oksendal2000}.  This strategy has similarities to 
``exponential time differencing''  or ``exact linear part'' 
numerical methods~\citep{eh:ltecn,bga:ltsmo,trefethen2005}.

In stochastic differential notation, the fluid equations with 
both thermal and particle forces can be expressed as
\begin{eqnarray}
\label{equ_stokes_DFT_thermal_particle}
d\hat{\mathbf{u}}_{\mathbf{k}} 
& = & 
-\alpha_{\mathbf{k}} \hat{\mathbf{u}}_{\mathbf{k}} 
dt 
+ \rho^{-1}\wp_{\mathbf{k}}^{\perp}\hat{\mathbf{f}}_{\mathbf{k}}dt
+ \sqrt{2D_{\mathbf{k}}}
\wp_{\mathbf{k}}^{\perp}d\mathbf{\tilde{B}}_{\mathbf{k}}(t),
\end{eqnarray}
where to simplify the notation 
the subscript will be dropped for the Fourier modes of the 
particle force density
so that 
$\mathbf{\hat{f}}_{\mathbf{k}}$ always refers to 
$\mathbf{\hat{f}}_{\mbox{\small prt},\mathbf{k}}$.

Approximating the particle force as constant over the time interval 
$ [t^{\prime},t] $ gives
\begin{eqnarray}
\label{analytic_sol_u}
\hat{\mathbf{u}}_{\mathbf{k}}(t)
& = & 
e^{-\alpha_{\mathbf{k}}{(t-t^{\prime})}}\hat{\mathbf{u}}_{\mathbf{k}}(t^{\prime})
+ \frac{1}{\rho \alpha_{\mathbf{k}}} 
\left(1 - e^{-\alpha_{\mathbf{k}}(t-t^{\prime})}\right)
\wp_{\mathbf{k}}^{\perp}\mathbf{\hat{f}}_{\mathbf{k}} (t^{\prime})  \\
\nonumber
& + & \sqrt{2D_{\mathbf{k}}} \int_{t^{\prime}}^{t} e^{-\alpha_{\mathbf{k}}(t - s)} 
\wp_{\mathbf{k}}^{\perp}d\mathbf{\tilde{B}}_{\mathbf{k}}(s), 
\end{eqnarray}
where $\wp_{\mathbf{k}}^{\perp}$ is the projection operation defined in 
\ref{equ_def_proj_perp} and 
$\int \cdot \wp_{\mathbf{k}}^{\perp}d\mathbf{\tilde{B}}_{\mathbf{k}}(s)$ 
denotes integration in the sense of Ito 
\citep{oksendal2000} 
over the 
projected complex-valued Brownian motion
$\mathbf{\tilde{B}}_{\mathbf{k}}(t)$ defined in Section 
\ref{section_thermal_forcing}. 

To obtain a numerical scheme for the fluid with finite time step $ \Delta t $, 
each mode is updated at discrete times $ n \Delta t $ using the analytic solution
\ref{analytic_sol_u}, yielding the stochastic recurrence equation
\begin{eqnarray}
\label{stoch_reccurence}
\hat{\mathbf{u}}_{\mathbf{k}}^{n + 1}
& = & 
e^{-\alpha_{\mathbf{k}}\Delta{t}}\hat{\mathbf{u}}_{\mathbf{k}}^{n}
+ \frac{1}{\rho \alpha_{\mathbf{k}}} 
\left(1 - e^{-\alpha_{\mathbf{k}}\Delta{t}}\right)\wp_{\mathbf{k}}^{\perp}
\hat{\mathbf{f}}_{\mathbf{k}}^{n} 
+ \wp_{\mathbf{k}}^{\perp}
\hat{\boldsymbol{\Xi}}_{\mathbf{k}}^n, 
\end{eqnarray}
where 
$\hat{\mathbf{u}}_{\mathbf{k}}^n = \hat{\mathbf{u}}_{\mathbf{k}}(n\Delta{t})$,
$\hat{\mathbf{f}}_{\mathbf{k}}^{n}  = \hat{\mathbf{f}}_{\mathbf{k}}(n\Delta{t})$,
and $\hat{\boldsymbol{\Xi}}_{\mathbf{k}}^n = 
\sigma_{\mathbf{k}}\boldsymbol{\tilde{\eta}}_{\mathbf{k}}$.

The notation $\boldsymbol{\tilde{\eta}}_{\mathbf{k}}$ denotes a three 
dimensional complex-valued random variable, with each real and imaginary 
component being an independent Gaussian random variable with mean
0 and variance 1.
The random variable $\hat{\boldsymbol{\Xi}}_{\mathbf{k}}^n$ 
accounts for the contributions of the stochastic integral 
in \ref{analytic_sol_u} over the time step.  The variance 
$\sigma_{\mathbf{k}}^2$ can be determined by 
Ito's Isometry \citep{oksendal2000} and is given by
\begin{eqnarray}
\sigma_{\mathbf{k}}^2 = 
\frac{D_{\mathbf{k}}}{\alpha_{\mathbf{k}}}
\left(1 - e^{-2\alpha_{\mathbf{k}}\Delta{t}}\right).
\end{eqnarray}

The constraint \ref{equ_DFT_real_constr} that ensures the real-valuedness 
of the velocity field is respected by only applying the 
update \ref{stoch_reccurence} 
to one 
member of each complex-conjugate pair, and then setting the new value for the 
partner mode as the complex conjugate of the computed mode.  
The condition \ref{equ_DFT_real_constr} also requires
that the modes $\hat{\mathbf{u}}_{k} $ with indices 
$ \mathbf{k} \in \mathcal{K} $ have zero 
imaginary part; this is enforced explicitly in each time step.

\subsubsection{Numerical Method for the Immersed Structures}
\label{section_numerical_method_particle}
A time-discretization for the equation 
(\ref{semidiscretization_particle_equ1}) is developed
for the advection of the elementary particles by 
integrating the fluid velocity field over a time step
and then averaging the integrated velocity over a spatial
neighborhood centered on the old particle position: 
\begin{eqnarray}
\label{num_method_particle_conv}
\mathbf{X}^{{n+1},[j]} - \mathbf{X}^{{n},[j]} & = & \sum_{\mathbf{m}} 
\delta_a(\mathbf{x}_{\mathbf{m}} - \mathbf{X}^{{n},[j]})
\int_{t_n}^{t_{n+1}} \mathbf{u}_{\mathbf{m}}(s) ds \Delta{x}^3, 
\end{eqnarray}
where $t_n = n\Delta{t}$
and $ \mathbf{X}^{{n},[j]} = \mathbf{X}^{[j]} (n \Delta t) $.

A precise integration of the fluid 
velocity $ \mathbf{u} $ is taken
which allows for time steps which underresolve the dynamics of some 
of the Fourier modes of the fluid.  This capability is important 
due to the wide range of time scales that may be associated with
the fluid modes and immersed structures in applications.   If  
the time integral is approximated through 
numerical methods built from (stochastic) Taylor expansions about 
discrete times, such as Runge-Kutta methods and their stochastic 
variations 
\citep{platen1992,dt:sspa},
then it is important that the 
method sufficiently resolve the fluctuations of the processes 
to capture cancellations that occur over time.  For 
instance, if the cancellation is not adequately captured, 
the numerical value of 
the integral of velocity
will be larger in magnitude than the 
actual time integrated velocity.  For immersed particles, 
this leads to an overly diffuse behavior where the particles
overshoot their correct positions each time step.

From \ref{equ_stokes_DFT_thermal2} the time scale
associated with the dynamics of the $\mathbf{k}^{th}$ mode 
of the fluid is ${1}/{\alpha_{\mathbf{k}}}$.  For the fastest modes 
of the fluid relevant for the immersed particle dynamics, 
the above considerations 
would place a severe restriction on the time step.
While 
there may be clever numerical methods involving (stochastic) 
Taylor expansions which perform better than anticipated, 
a different approach will be taken here.

To develop a method that remains accurate for a range of
time steps, from those that fully resolve, partially resolve,
or completely underresolve the fluid modes, we calculate the time integral in
\ref{num_method_particle_conv} by substitution of the analytical 
expression \ref{analytic_sol_u} for the Fourier
modes of the fluid velocity field.   We recall that this approximation
only assumes that the elementary 
particle positions and forces can be considered
frozen over a time step. The resulting numerical scheme can therefore
be expected to be accurate provided the time step $ \Delta t $ is
chosen small compared to the time scales of the immersed structures,
but with no restriction on the size of the time step relative to the
time scales of the fluid modes.  We will explain this property
more precisely through numerical error analysis in 
Section~\ref{section_accuracy}.

In updating the elementary 
particle positions in the numerical method,
the time integral in \ref{num_method_particle_conv} will be 
simulated as a random variable
\begin{eqnarray}
\mathbf{\Gamma}_{\mathbf{m}}^n = \int_{t_n}^{t_{n+1}} 
\mathbf{u}_{\mathbf{m}}(s) ds 
= \sum_{\mathbf{k}} \hat{\mathbf{\Gamma}}_{\mathbf{k}}^n
\exp\left({{i2\pi{\mathbf{k}}\cdot \mathbf{m}}/{N}}\right), 
\end{eqnarray}
\begin{eqnarray}
\hat{\mathbf{\Gamma}}_{\mathbf{k}}^n
=
\int_{t_n}^{t_{n+1}} \hat{\mathbf{u}}_{\mathbf{k}}(s) ds, 
\end{eqnarray}
with $ \hat{\mathbf{u}}_{k} (s) $ given by (\ref{analytic_sol_u}).

By using standard techniques from stochastic calculus,
the time integral can be evaluated by 
defining $ \hat{\mathbf{\Gamma}}_{\mathbf{k}}^n $ to give 
the Gaussian random variable
\begin{eqnarray}
\label{equ_def_lambda_k_1}
&& \\
\nonumber
\hat{\mathbf{\Gamma}}_{\mathbf{k}}^n
& = & 
-\frac{e^{-\alpha_{\mathbf{k}}\Delta{t}} - 1}{\alpha_{\mathbf{k}}}
\hat{\mathbf{u}}_{\mathbf{k}}^n
+ 
\left(
\frac{\Delta{t}}{\alpha_{\mathbf{k}}}
+ 
\left(\frac{1}{\alpha_{\mathbf{k}}}\right)^2
\left(
e^{-\alpha_{\mathbf{k}}\Delta{t}} - 1
\right)
\right)
\rho^{-1}
\wp_{\mathbf{k}}^{\perp}
\hat{\mathbf{f}}_{\mathbf{k}}^n \\
\nonumber
& - & 
\frac{\sqrt{2D_{\mathbf{k}}}}{\alpha_{\mathbf{k}}}
\int_{t_n}^{t_{n+1}} e^{-\alpha_{\mathbf{k}}(t_{n+1} - r)} 
\wp_{\mathbf{k}}^{\perp}
d\mathbf{\tilde{B}}_{\mathbf{k}}(r)
+ \frac{\sqrt{2D_{\mathbf{k}}}}{\alpha_{\mathbf{k}}}
\left(
\wp_{\mathbf{k}}^{\perp}
\mathbf{\tilde{B}}_{\mathbf{k}}(t_{n+1})
- 
\wp_{\mathbf{k}}^{\perp}
\mathbf{\tilde{B}}_{\mathbf{k}}(t_n)
\right).
\end{eqnarray}
Using \ref{analytic_sol_u} at times $n\Delta{t}$ and $ (n+1)\Delta{t} $,
this can be expressed more simply as 
\begin{eqnarray}
\label{equ_def_lambda_k_2}
\mathbf{\hat{\Gamma}}_{\mathbf{k}}^n & = & -\frac{1}{\alpha_{\mathbf{k}}}
\left(\hat{\mathbf{u}}_{\mathbf{k}}^{n+1} 
- \hat{\mathbf{u}}_{\mathbf{k}}^n\right) 
 +  
\rho^{-1}
\frac{\wp_{\mathbf{k}}^{\perp} 
\hat{\mathbf{f}}_{\mathbf{k}}^n}
{\alpha_{\mathbf{k}}}\Delta{t}  \\
\nonumber
& + &  
\frac{\sqrt{2D_{\mathbf{k}}}}{\alpha_{\mathbf{k}}} 
\left(
\wp_{\mathbf{k}}^{\perp}
\mathbf{\tilde{B}}_{\mathbf{k}}(t_{n+1})
- 
\wp_{\mathbf{k}}^{\perp}
\mathbf{\tilde{B}}_{\mathbf{k}}(t_n)
\right).
\end{eqnarray}

The numerical scheme to update the elementary 
particle positions 
is then given by 
\begin{eqnarray}
\label{equ_particle_numerical_method}
\mathbf{X}^{{n+1},[j]} - \mathbf{X}^{{n},[j]} & = & \sum_{\mathbf{m}} 
\delta_a(\mathbf{x}_{\mathbf{m}} - \mathbf{X}^{{n},[j]})
\mathbf{\Gamma}_{\mathbf{m}}^n\Delta{x}^3,
\end{eqnarray}
where $\mathbf{\Gamma}_m^n$ is generated each time step.  To consistently 
update the particle positions with the velocity field of the fluid 
it is required that $\hat{\mathbf{\Gamma}}_{\mathbf{k}}^n$ be generated 
with the correct 
correlations to the fluid modes at the beginning and end of each 
time step, $\{\mathbf{u}_{\mathbf{k}}^{n}\}$ and 
$\{\mathbf{u}_{\mathbf{k}}^{n + 1}\}$.
In the next Subsection, a practical approach for doing so is presented.

\subsubsection{Method for Generating Modes of the Time 
Integrated Velocity Field}
\label{section_method_consistently_gen_modes}
Since the modes $\hat{\mathbf{\Gamma}}_{\mathbf{k}}^n$ of the time 
integrated velocity field and the modes $\hat{\mathbf{u}}_{\mathbf{k}}^{n}$
and  $\hat{\mathbf{u}}_{\mathbf{k}}^{n+1}$ of the velocity field evaluated at
the beginning and end of a time step
are not statistically independent, some care must be 
taken in generating the corresponding random variables that are
used in the simulation.  Since
 $\hat{\mathbf{\Gamma}}_{\mathbf{k}}^n$,  
 $\hat{\mathbf{u}}_{\mathbf{k}}^{n}$, and
$\hat{\mathbf{u}}_{\mathbf{k}}^{n+1}$ are 
jointly Gaussian distributed random variables with mean zero,
we need only ensure they have 
the correct covariances between their components.    Since the 
real and imaginary parts of each mode are independent, 
we shall for clarity consider only the real components
with the understanding that the imaginary components are handled 
in a similar manner.  

In deriving a method to generate the time integrated 
field, it is useful to express the real part
$\mbox{Re}(\mathbf{\hat{\Gamma}}_{\mathbf{k}}^n)$ 
in terms of the following random variables 
\begin{eqnarray}
\mbox{Re}(\mathbf{\hat{\Gamma}}_{\mathbf{k}}^n) 
   & = & 
\wp_{\mathbf{k}}^{\perp}
\mathbf{A}_0 + 
\wp_{\mathbf{k}}^{\perp}
\mathbf{A}_1 + 
\wp_{\mathbf{k}}^{\perp}
\mathbf{A}_2,
\end{eqnarray} 
with
\begin{eqnarray}
\label{equ_def_Z_0}
\mathbf{A}_0  & = & 
\frac{1 - e^{-\alpha_{\mathbf{k}}\Delta{t}}}{\alpha_{\mathbf{k}}}
      \mbox{Re}(\hat{\mathbf{u}}_{\mathbf{k}}^n) 
+ 
\left(
\frac{\Delta{t}}{\alpha_{\mathbf{k}}} 
+ 
\left(\frac{1}{\alpha_{\mathbf{k}}}\right)^2
\left(
e^{-\alpha_{\mathbf{k}}\Delta{t}} - 1
\right)
\right)
\rho^{-1}
\mbox{Re}(
%\wp_{\mathbf{k}}^{\perp}
\hat{\mathbf{f}}_{\mathbf{k}}^n) \\
\label{equ_def_G_1}
\mathbf{A}_1            
& = & -\frac{\sqrt{2D_{\mathbf{k}}}}{\alpha_{\mathbf{k}}}\int_{t_n}^{t_{n+1}} 
                              e^{-\alpha_{\mathbf{k}}(t_{n+1} - s)}
                      \mbox{Re}(d\mathbf{\tilde{B}}_{\mathbf{k}}(s)) \\
\nonumber
          & = & -\frac{1}{\alpha_{\mathbf{k}}}
\boldsymbol{\Xi}_{\mathbf{k}}^n \\
\label{equ_def_G_2}
\mathbf{A}_2     & = & \frac{\sqrt{2D_{\mathbf{k}}}}{\alpha_{\mathbf{k}}} 
               \mbox{Re}\left(\mathbf{\tilde{B}}_{\mathbf{k}}(t_{n+1})
                              - \mathbf{\tilde{B}}_{\mathbf{k}}(t_n)\right) \\
\nonumber
        & = & \frac{\sqrt{2D_{\mathbf{k}}}}{\alpha_{\mathbf{k}}} \int_{t_n}^{t_{n+1}} 
                              \mbox{Re}(d\mathbf{\tilde{B}}_{\mathbf{k}}(s)). 
\end{eqnarray}
The 
random variables were obtain by reorganizing the terms of 
\ref{equ_def_lambda_k_1}.

This expression recasts the problem of determining the correlations of
$\mbox{Re}(\mathbf{\hat{\Gamma}}_{\mathbf{k}}^n)$ 
to the problem of determining 
the correlations of $\mathbf{A}_0$, $\mathbf{A}_1$ and $\mathbf{A}_2$ with
each other and the modes of the fluid velocity.  A convenient feature of 
this approach is that $\mathbf{A}_0$ is already determined at the beginning 
of the time step, and is statistically independent of 
$\mathbf{A}_1$ and $\mathbf{A}_2$ by the independent increment property of 
Brownian motion.  This reduces the problem to finding the covariance
 of $\mathbf{A}_1$ and $\mathbf{A}_2$.  A useful
identity for Ito integrals in this context is~\citep{oksendal2000} 
\begin{eqnarray}
\label{equ_ito_inner_prod}
E\left(\int_0^{t} f(s) dB_s \int_0^{t} g(r) dB_r\right)
= \int_0^{t} f(s)g(s) ds, 
\end{eqnarray}
where the notation $E(\cdot)$ 
denotes expectation with respect the underlying 
Brownian motion \citep{oksendal2000}.

Using \ref{equ_ito_inner_prod}, the covariance is given by 
\begin{eqnarray}
E(\mathbf{A}_1^{(j)} \mathbf{A}_2^{(j)})
& = & -\frac{2D_{\mathbf{k}}}{\alpha_{\mathbf{k}}^3}
\left(1 - \exp\left(-\alpha_{\mathbf{k}}\Delta{t}\right) \right),
\end{eqnarray}
where the parenthesized superscripts denote the indices 
of the vector components.  When $j \not= j^{\prime}$ the components 
$\mathbf{A}_1^{(j)}$ and $\mathbf{A}_2^{(j^{\prime})}$ are 
independent and have zero correlation.

The variance of the components of $\mathbf{A}_1$ and $\mathbf{A}_2$ 
are given by 
\begin{eqnarray}
E(|\mathbf{A}_1^{(j)}|^2) & = & \frac{D_{\mathbf{k}}}{\alpha_{\mathbf{k}}^3}
                \left(1 - \exp\left(-2\alpha_{\mathbf{k}}\Delta{t}\right) \right) \\
E(|\mathbf{A}_2^{(j)}|^2) & = & \frac{2D_{\mathbf{k}}}{\alpha_{\mathbf{k}}^2}\Delta{t}. 
\end{eqnarray}

From the numerical updating of the fluid variables described in 
Subsection~\ref{section_numerical_method_fluid}, 
$\mathbf{A}_1 = -\frac{1}{\alpha_{\mathbf{k}}}
\boldsymbol{\Xi}^n_{\mathbf{k}}$
is already known each time step, so only $\mathbf{A}_2$ need be
generated.  Obtaining this random variable with the correct correlations
can be accomplished by generating new 
standard Gaussian random variables 
$\boldsymbol{\eta}^{(j)}$ 
(independent in $j$ with mean 0 and variance 1) 
and by taking the linear combination of the two random variables
$\mathbf{A}_1^{(j)}$
 and $\boldsymbol{\eta}^{(j)}$ 
 given by
\begin{eqnarray}
\label{equ_gaussian_rep_1}
\mathbf{A}_2^{(j)} & = & 
a_1 \mathbf{A}^{(j)}_1 + a_2 \boldsymbol{\eta}^{(j)},
\end{eqnarray}
with 
\begin{eqnarray}
a_1  = 
\frac{E\left(\mathbf{A}_1^{(j)}\mathbf{A}_2^{(j)}\right)}
{E\left(|\mathbf{A}_1^{(j)}|^2\right)}
\end{eqnarray}
and
\begin{eqnarray}
a_2 & = & \sqrt{ 
          \frac{E\left(|\mathbf{A}_1^{(j)}|^2\right)
E\left(|\mathbf{A}_2^{(j)}|^2\right)
             - E\left(\mathbf{A}_1^{(j)}\mathbf{A}_2^{(j)}\right)^2}
             {E\left(|\mathbf{A}_1^{(j)}|^2\right)}}.
\end{eqnarray}

In this manner, $\mbox{Re}(\boldsymbol{\hat{\Gamma}}_{\mathbf{k}}^{n})$ can be
generated from $\mathbf{A}_0$, $\mathbf{A}_1$, and
$\mathbf{A}_2$ with proper accounting of correlations with the modes of
the velocity field.  The imaginary component
$\mbox{Im}(\boldsymbol{\hat{\Gamma}}_{\mathbf{k}}^n)$ is
generated in an analogous manner.

\section{Accuracy of the Method}
\label{section_accuracy}
In this section the accuracy of the numerical method is
investigated.  Three asymptotic scaling regimes of the time step
are considered.  The first regime applies when the time step is 
taken sufficiently small to fully resolve the dynamics 
of the fluid.  The second applies when the time step is taken large 
and completely underresolves the dynamics of  the fluid.  
We finally consider the case in which the time step 
resolves some but not all of the fluid modes.

Formal error estimates are given which show how the numerical errors 
scale with respect to the time step and various key parameters.  
While a rigorous analysis making use of standard stochastic Taylor 
expansion approaches 
\citep{platen1992} can be carried out for time steps which are small when 
compared to the time scales of the fluid and immersed structure 
dynamics, a completely rigorous analysis of the numerical error 
when the time step is large and underresolves a subset of the 
fluid modes is considerably more difficult.

An important feature 
of the numerical method is the way in which the statistical 
contributions of the fluid dynamics are taken into account, even when the 
fluid dynamics are underresolved.  As discussed in Subsections
\ref{section_numerical_method_fluid} and
\ref{section_numerical_method_particle},
the random increments of the elementary 
particle positions and fluid modes are 
simulated in such a way that the correct statistics and correlations
are preserved over time steps 
 which need only be small compared to the 
time scales of the immersed structures. 
  While the time step relative
to the time scale of the fastest modes of the fluid may be large, this procedure
helps keep the local time discretization error small.
By contrast, standard finite difference schemes would
generally have poor accuracy once the 
time step exceeded the time scales of the fastest fluid modes.

To quantify the accuracy of the method, 
the strong error is considered, as defined in \citep{platen1992}.
Let
$\mathbf{X}^{[j]}(t)$ 
denote 
the exact solution of equation 
\ref{semidiscretization_particle_equ1}
for the 
elementary 
 particles and
$\hat{\mathbf{u}}_{\mathbf{k}}(t)$ 
denote the exact solution to
equation
\ref{semidiscretization_fluid_equ1}
for the Fourier modes of the 
velocity field of the fluid.
Let the numerically computed trajectories of 
the elementary 
particles be denoted by 
$\mathbf{\tilde{X}}^{[j]}(t)$ 
and 
the numerically 
computed fluid modes be denoted by 
$\mathbf{\tilde{\hat{u}}}_{\mathbf{k}}(t)$. 
Since we shall be interested in the error associated 
with a typical elementary 
particle, the superscript $j$ will
be dropped throughout the discussion.

The strong error of the numerical method 
associated with the $\mathbf{k}^{th}$ mode of the fluid is defined as
\begin{eqnarray}
\hat{e}_{\mbox{\small fld},\mathbf{k}}(\Delta{t}) & = & 
E\left(
\left|
\hat{\mathbf{u}}_{\mathbf{k}}(\Delta{t})
-
\tilde{\hat{\mathbf{u}}}_{\mathbf{k}}(\Delta{t})
\right|
\right),
\end{eqnarray}
and the strong error associated with an elementary 
particle is defined as
\begin{eqnarray}
e_{\mbox{\small prt}}(\Delta{t}) & = & 
E\left(
\left|
\mathbf{X}(\Delta{t}) - 
\mathbf{\tilde{X}}(\Delta{t})
\right|
\right). 
\end{eqnarray}

The error associated to the velocity field in physical space is
defined as
\begin{eqnarray}
e_{\mbox{\small fld}}(\Delta{t}) & = & 
E\left(
\frac{1}{L^3}
\sum_{\mathbf{m}}
\left|
\mathbf{u}_{\mathbf{m}}(\Delta{t})
-
\tilde{\mathbf{u}}_{\mathbf{m}}(\Delta{t})
\right|
\Delta{x}^3
\right). 
\end{eqnarray}

For further discussion of the strong error see \citep{platen1992}.

The error expressions above and the estimates given below are 
intended to characterize the ``typical'' error for 
the numerical method; in reality they will of course depend 
on the particular configuration the elementary 
particles happen to be 
in at the beginning of a time step and the details of the 
forces acting between them.  For the purposes of 
describing the errors incurred in the numerical method's 
handling of the force interactions, we shall therefore concern ourselves 
with describing how the errors scale with respect to the various numerical 
parameters.  

In the derivation of the estimates we 
quantify the error incurred by the numerical method's 
representation of the stochastic (thermal) components 
of the structural 
and fluid dynamics.  The estimates presented
follow from a systematic formal analysis of the errors resulting 
from the discretization of the stochastic and deterministic components 
of the dynamics, 
including their interaction during a time step.  This calculation leads to a 
uniformly valid expression for time steps sufficiently small that the 
elementary 
 particles do not move appreciably (relative to their size) during a 
time step; no assumption is made in the derivation about the  
magnitude of the time step relative to the time scales of the fluid modes.
As the resulting derivations are somewhat technical, we shall  
in the present paper be content to state the error estimates,
discuss their significance, and confirm their validity in a few special
cases by numerical simulation.    For 
 a detailed derivation see 
~\citep{atzberger2005error}.

\subsection{Error Estimates for Time Steps
 which Fully Resolve the Fluid Dynamics}
\label{section_error_estimate_small_time}
When the time step is taken sufficiently small
so that the dynamics of all modes of the fluid 
are resolved by the stochastic immersed boundary
method ($\Delta{t} \ll \min\frac{1}{\alpha_{\mathbf{k}}}$), 
the following error estimates can be established: 
\begin{eqnarray}
\hat{e}_{\mbox{\small fld},\mathbf{k}}(\Delta{t})
& \approx & 
\frac{M F^*}{\rho}\delta_{a,\mathbf{k}}^*
\left(
\frac{M}{\ell_F}
+ 
C
\frac{1}{a}
\right)
\left(
C' v_{\mbox{\small frc}} 
+ 
C'' v_{\mbox{\small thm}} 
\right)
\Delta{t}^2
\end{eqnarray}

\begin{eqnarray}
e_{\mbox{\small fld}}(\Delta{t})
& \approx & 
\frac{M F^*}{\rho a^{3/2}L^{3/2}}
\left(
\frac{M}{\ell_F}
+ 
C
\frac{1}{a}
\right)
\left(
C' v_{\mbox{\small frc}} 
+ 
C'' v_{\mbox{\small thm}} 
\right)
\Delta{t}^2
\end{eqnarray}

\begin{eqnarray}
\label{equ_error_e_prt_small_time_step}  
e_{\mbox{\small prt}}(\Delta{t}) 
& \approx &
\left(
Q_1
v_{\mbox{\small thm}}^2 
+
C v_{\mbox{\small frc}} 
v_{\mbox{\small thm}} 
+
C'
v_{\mbox{\small frc}}^2 
\right)
\frac{\Delta{t}^2}{a} \\
\nonumber
& + &
\frac{M F^*}{\rho a^3}
\left(
\frac{M}{\ell_F}
+ 
C''
\frac{1}{a}
\right)
\left(
C''' v_{\mbox{\small frc}} 
+ 
C'''' v_{\mbox{\small thm}} 
\right)
\Delta{t}^3, 
\end{eqnarray}
where 
$M$ is the number of elementary 
 particles,
$F^*$ is the magnitude of the force acting on the 
elementary 
 particles, and $\ell_F$ is the length scale 
associated with changes in the particle force of 
order $F^*$.  It will be assumed throughout that
$ a \lesssim \ell_{F} $.
The factor $\hat{\delta}_{a,\mathbf{k}}^*$
is the magnitude of the  Fourier coefficient
for mode $\mathbf{k}$
of the function $\delta_a$, averaged over all shifts
(see Appendix \ref{appendix_delta_func_hat}).
  We remark that
$ \hat{\delta}_{a,\mathbf{k}}^{*} \approx 1/L^{3} $ 
for $ |\mathbf{k}| \ll L/a $, while
$ \hat{\delta}_{a,\mathbf{k}}^{*} $ decays rapidly 
for $ |\mathbf{k}| \gg L/a $.

In this notation the factors $C$ which are superscripted with primes
are approximately independent of the physical 
parameters, and can be thought of as order unity constants.  To avoid 
cumbersome notation and overly emphasizing the role of these factors 
the notation is reused in each equation, with the understanding that 
$C$ denotes distinct factors for each estimate.  The subscripted factors 
$Q$ are also approximately independent of the physical parameters.  They 
are distinguished since numerical values will be estimated for these factors 
in order to make a comparison between the theoretical estimates
and numerical simulations in the case that $F^* = 0$.

To simplify the expressions the following terms 
are defined $v_{\mbox{\small thm}} = \sqrt{{k_B{T}}/{\rho a^3}}$
and $v_{\mbox{\small frc}} = {F^*}/{\mu a}$.  The factor
$v_{\mbox{\small thm}}$ can be interpreted via the 
equipartition theorem of statistical mechanics 
\citep{IBler:mcsp,IBrk:sp2} 
as the velocity scale of thermal fluctuations of an 
elementary particle of size 
$ a $, since the associated mass will be proportional to $ \rho a^{3}$.
The term
$v_{\mbox{\small frc}}$ can be interpreted
as the velocity scale associated with the motion of a particle 
of size $a$ in a viscous fluid when a force of magnitude $F^*$ 
is applied to the particle, since the friction coefficient of a particle is 
generally proportional to $ \mu a $ 
\citep{jfb:sd}.  

The error estimates
indicate that the 
stochastic immersed boundary method has strong first 
order accuracy as the time step is taken small.  
An error proportional 
to $ \Delta t^{3} $ is included in 
$e_{\mbox{\small prt}}(\Delta{t})$ because its coefficient in certain
circumstances can make it comparable to the
$ \Delta t^{2} $.
We remark that the reported proportionality of the errors 
with respect to $ M$, the number of elementary particles, is based on 
a worst-case scenario where all $ M$ particles are clustered near each other.
In general the error is expected to scale with a smaller factor
reflecting the actual number of particles clustered in a region.
Since this depends on details of the force interaction between 
particles,  obtaining a more precise error estimate is technically 
involved and somewhat application dependent.
While in practice the actual numerical error will likely be 
somewhat better than these factors indicate, we leave further 
refinements to future work in the context of specific
applications.

An important observation is that in the absence of forces on 
the immersed structures ($F^{*}=0$), 
the fluid modes are simulated exactly (for the reasons discussed in 
Subsection \ref{section_numerical_method_fluid}).  Only the elementary 
particle 
dynamics incur  
a temporal discretization error in this case (see Subsection 
\ref{section_numerical_method_particle}), with the strong error incurred 
being of first order.  A more conventional 
time 
stepping scheme based on
finite differences 
would typically
incur an error 
for the 
velocity mode 
$ \hat{\mathbf{u}}_{\mathbf{k}} $ which includes a 
contribution which scales as 
$ C N^{-3/2}v_{\mbox{\small thm}} 
(\alpha_{\mathbf{k}} \Delta t)^{n+1/2}  $ 
for some integer $ n $.   Such an error fails to remain small compared to 
the actual velocity change over a time step as soon as 
$\Delta{t} \gtrsim 1/\alpha_{\mathbf{k}}$.
For the numerical method developed in Section \ref{section_numerical_method},
the exact representation of the stochastic fluid dynamics, apart from the 
response to the forces exerted by the immersed structures, maintains better 
accuracy even as the time step underresolves the fluid dynamics.

\begin{figure}[h*]
\centering
\epsfxsize = 4in
\epsffile[98 233 490 558]{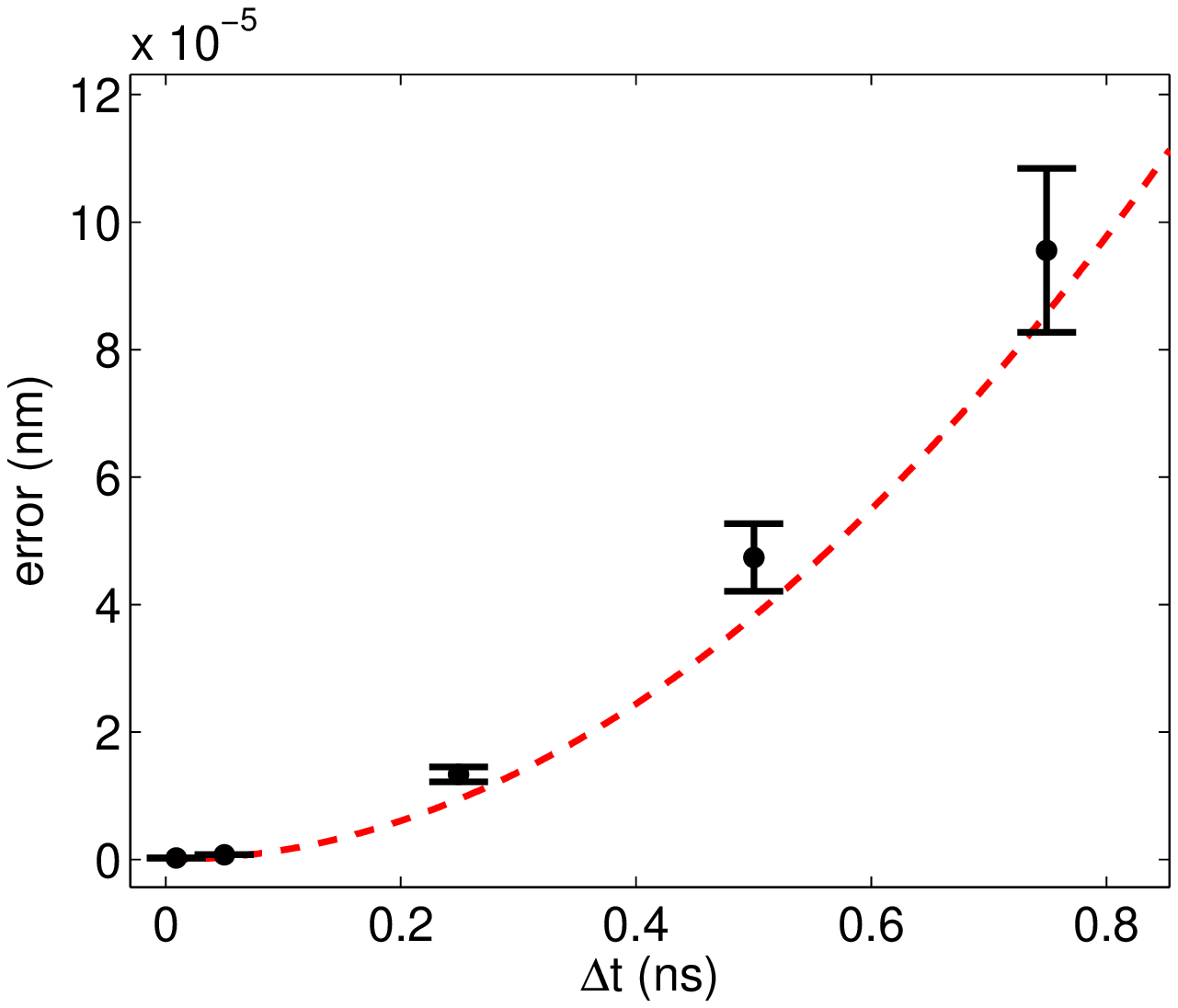}
\caption[Comparison of the numerical error and 
non-rigorous analytic estimate for small and intermediate time steps]
{A comparison of the analytic error estimate for the 
elementary particle positions 
 $e_{\mbox{\tiny prt}}(\Delta{t})$ 
given in equation \ref{equ_error_e_prt_small_time_step}
with a numerical estimation of the error  
for the small to intermediate time step regime 
$\Delta{t} \ll \rho a^2/\mu$
with $F^* = 0$ and parameter values given in 
Table 
\ref{table_params_diffusion_compare}.
The dashed line denotes $e_{\mbox{\tiny prt}}(\Delta{t})$.
The data points denote the error as estimated from
numerical simulations using the method.  The error 
bars indicate one standard deviation of the sampled values.  
To compute the numerical estimates of the error,
simulations were performed with each given time step
$\Delta{t}$ and compared with an ensemble of reference 
trajectories obtained from a high resolution simulation 
sufficient to resolve the dynamics of all $N^3$ modes of the fluid.
The high resolution simulation had a time step of
$\Delta{t} = 10^{-3}\mbox{ns} < 
\min 1/\alpha_{\mathbf{k}}$, where 
$\min 1/\alpha_{\mathbf{k}} = 3.94\times 10^{-2} \mbox{ns}$
for the parameter values given in Table 
\ref{table_params_diffusion_compare}.
From the results, we see that
the estimate given in equation \ref{equ_error_e_prt_small_time_step} 
quantifies the error well for $\Delta{t} \ll 
\rho a^2/\mu = 0.976 \mbox{ns}$.}
\label{figure_nonrigorous_error_e_prt_small_time}
\end{figure}

As demonstrated in Figure 
\ref{figure_nonrigorous_error_e_prt_small_time},
the theoretical error estimate for the elementary 
particles agrees well with numerical simulations
in the absence of particle forces $(F^* = 0)$.
The numerical results were obtained from simulations
of the fluid-particle system
with physical parameters in Table 
\ref{table_params_diffusion_compare}.
  In the comparison,
the factors $Q$ were computed from theoretical expressions
arising in the derivation of the estimates, and their
numerical values are given in Appendix 
\ref{appendix_constants_accuracy_error}.
It should be emphasized that the error estimates are stated as formal 
approximations, not as upper bounds.  In the case that there are 
particle forces, the discretization error depends on a number of 
details of the force structure, and therefore 
 numerical comparison 
with simulations is left to future work in the context of specific 
applications.

\subsection{Error Estimates for Time Steps
which Underresolve All Fluid Modes}
\label{section_error_estimate_large_time}
We now present estimates for the error 
of the numerical method when the time step 
is taken large enough to underresolve all modes of the fluid,
but always small enough to resolve the elementary 
particle dynamics: 
\begin{equation}
\max \frac{1}{\alpha_{\mathbf{k}}} \ll \Delta{t} \ll 
\tau_{\mbox{\small mov}}(a).
\label{equ_time_step_large}
\end{equation}
The notation 
$\tau_{\mbox{\small mov}}(a)$ denotes the time
required for an elementary 
 particle to move a displacement equal
to its size $a$ either by advection or diffusion.  
In this regime the following error estimates can be established: 
\begin{eqnarray}
&& \label{equ_error_e_fld_large_time_step}  \\
\nonumber
\hat{e}_{\mbox{\small fld},\mathbf{k}}(\Delta{t})
& \approx & 
\frac{M F^* L^2}{\mu |\mathbf{k}|^2}
\delta_{a,\mathbf{k}}^*
\left(
\frac{M}{\ell_F}
+
%\frac{\Delta{x}^3}{a^4}
C
\frac{1}{a}
\right)
\left(
C'
v_{\mbox{\small frc}}\Delta{t}
+ 
C''
\sqrt{D}\Delta{t}^{{1}/{2}}
\right)
\end{eqnarray}

\begin{eqnarray}
&& \\
\nonumber
e_{\mbox{\small fld}}(\Delta{t})
& \approx &
\sqrt{\frac{a}{L}} 
\frac{MF^*}{\mu L}
\left(
\frac{M}{\ell_F}
+
C
\frac{1}{a}
\right)
\left(
C'
v_{\mbox{\small frc}}\Delta{t}
+ 
C''
\sqrt{D}\Delta{t}^{{1}/{2}}
\right)
\end{eqnarray}

\begin{eqnarray}
\label{equ_error_e_prt_large_time_step}
&& \\
\nonumber
e_{\mbox{\small prt}}(\Delta{t}) 
& \approx &
Q_2
\frac{D}{a}
\Delta{t}
\\
\nonumber
& + &
\left(
C
\frac{M}{\ell_F}
+
C'
\frac{1}{a}
\right)
\left(
\sqrt{D}
v_{\mbox{\small frc}}
\Delta{t}^{{3}/{2}} + v_{\mbox{\small frc}}^{2} \Delta{t}^{2}\right), 
\end{eqnarray}
where $D$ denotes the diffusion coefficient of an immersed particle
(see Section \ref{section_diffusion_coeff})
and the factors $ C $ and $Q$
denote order unity nondimensional constants as discussed in
Subsection~\ref{section_error_estimate_small_time}.
% and $\nu = \mu/\rho$ denotes the kinematic viscosity.  
The other terms are the same as in Subsection 
\ref{section_error_estimate_small_time}.

The smaller powers of $ \Delta t $ appearing in the error estimates may 
suggest that the accuracy is deteriorating more rapidly with respect 
to the time step in the underresolved regime under discussion, but in fact the opposite is 
true.  The error estimates reported above are in fact, for the range of time 
steps defining the underresolved regime, considerably smaller than the 
extrapolation of the error estimates in 
Subsection~\ref{section_error_estimate_small_time} which are 
valid only for the fully 
resolved regime.  Indeed, the ratio of terms appearing in the above 
estimates to  corresponding terms in the equations in Subsection~\ref{section_error_estimate_small_time}
involve ratios such as 
$ \rho L^{2}/(\mu |\mathbf{k}|^{2} \Delta t) $,
$ D^{1/2}/(v_{\mbox{\small thm}} \Delta{t}^{1/2}) $, 
and $ L^{1/2} a^{3/2} \rho/\mu \Delta t $, 
 all of which are much smaller 
than one in the asymptotic regime~\ref{equ_time_step_large}.

A more important point is that
the numerical errors remain small relative to the 
changes in the system variables throughout this range of time steps, 
so that the numerical method maintains accuracy for all
$ \Delta t \lesssim \tau_{\mbox{\small mov}}(a) $.  This can be 
seen by observing 
that the changes in the system variables over a time step falling in the 
regime~\ref{equ_time_step_large} 
can be estimated as 
\begin{eqnarray}
|\delta\hat{\mathbf{u}}_{\mathbf{k}}| & \approx & C 
\frac{M F^{*} \delta^{*}_{a,k}}{\rho \alpha_{\mathbf{k}}} 
+ C' \frac{v_{\mbox{\small thm}}}{N^{3/2}}, \\
|\delta \mathbf{u}| & \approx & C v_{\mbox{\small frc}} 
+ C' v_{\mbox{\small thm}}, \\
|\delta \mathbf{X}| & \approx & C\sqrt{D \Delta t} + C' v_{\mbox{\small frc}} \Delta t,
\label{equ_var_changes_large_time}
\end{eqnarray}
where the factors $ C $ 
denote order unity nondimensional constants as discussed in
Subsection~\ref{section_error_estimate_small_time}.
The notation $|\delta[\cdot]|$ indicates the absolute value of an increment 
of a variable over the time step.

Since the velocity field of the fluid
is completely underresolved, 
it changes by an amount 
comparable to its equilibrium value independently of the size of the time step.
The ratios
of the error estimates to the corresponding true changes in the system variables
in~\ref{equ_var_changes_large_time} can be bounded by sums and products of the
nondimensional groups $ \sqrt{D \Delta t}/a $,
$  v_{\mbox{\small frc}} \Delta t/a $, $ M \frac{a}{L} $,
and $ M \frac{a}{\ell_F} $.  The former two nondimensional groups involving the time step
are both small by definition
of the constraint $ \Delta t \ll \tau_{\mbox{\small mov}} (a) $ determining
the asymptotic regime~\ref{equ_time_step_large} under consideration.  The
nondimensional parameters 
 $ M \frac{a}{L} $ and $ M \frac{a}{\ell_{F}} $ 
 will be order unity or smaller 
 when the
system involves a small number of elementary particles.  When the system contains a large number of elementary particles, these nondimensional groups 
can become large and the error estimates become worse.  While it is certainly to be expected that the presence of more complex structures involving more elementary particles will generally incur more error in the numerical simulation, we stress that the scaling of our errors with large $ M$ are surely too pessimistic.  We therefore do not lay undue emphasis on the behavior of the errors for large   $ M $, which in any event will depend heavily on the details of the force structure.

We emphasize that unlike traditional numerical analysis the presence of terms 
proportional to $ \Delta t $ in the error estimate~\ref{equ_error_e_prt_large_time_step} 
does not imply that the method is inconsistent.  It must be remembered that these error 
estimates are appropriate not in the $ \Delta t \downarrow 0 $ limit, but rather in the 
asymptotic regime~\ref{equ_time_step_large}.  A more careful consideration of the 
sizes of the errors relative to the true changes in the system variables over a 
time step shows that our numerical method does in fact remain accurate for all 
time steps $ \Delta t \ll \tau_{\mbox{\small mov}} (a) $, even if the fluid modes are 
underresolved.  Vital to this result was the use of the stochastic integral 
formula~\ref{analytic_sol_u} for the action of the thermal forces on the velocity 
field of the fluid, 
and the systematic 
 consideration in 
Subsection~\ref{section_numerical_method_particle} of how to correlate the stochastic 
component of the velocity field of the fluid 
with the random motion of the immersed structures.
Without these developments, the resulting numerical method could not be expected to have
good accuracy for time steps in the regime~\ref{equ_time_step_large}.

\begin{figure}[h*]
\centering
\epsfxsize = 4in
\epsffile[95 232 486 541]{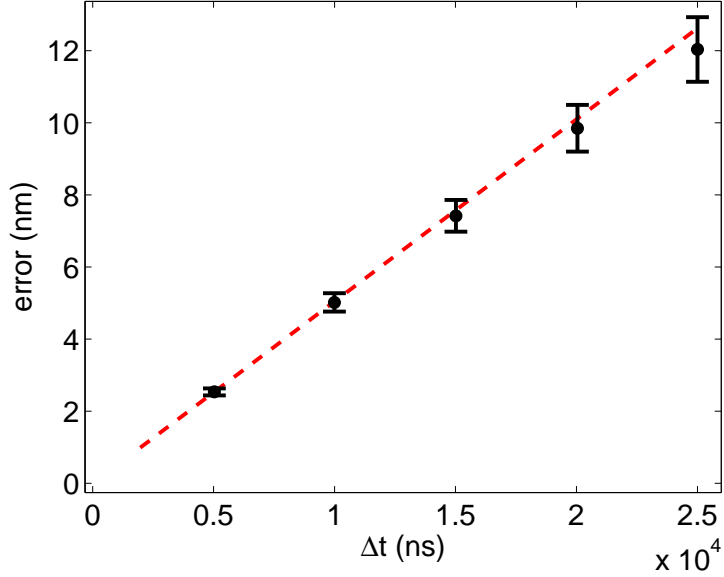}
\caption[Comparison of the numerical error and non-rigorous analytic
estimate for large time steps.]
{A comparison of the analytic error estimate for the 
elementary particle positions 
 $e_{\mbox{\tiny prt}}(\Delta{t})$ 
given in equation \ref{equ_error_e_prt_large_time_step}  
with a numerical estimation of the error for the large time step regime 
$\max{1}/{\alpha_{\mathbf{k}}} \ll
\Delta{t} \ll \tau_{\small mov}$
with 
 $F^* = 0$ and parameter values given in 
Table \ref{table_params_diffusion_compare}.
The dashed line denotes $e_{\mbox{\tiny prt}}(\Delta{t})$.
The data points denote the error as estimated from
numerical simulations using the method.  The error 
bars indicate one standard deviation of the sampled values.  
To compute the numerical estimates of the error,
simulations were performed with each given time step
$\Delta{t}$ and compared with an ensemble of reference 
trajectories obtained from a high resolution simulation 
with a time step of
$\Delta{t} = 1.0 \mbox{ns} < 
\max {1}/{\alpha_{\mathbf{k}}}$ where 
$\max 1/\alpha_{\mathbf{k}} = 25.4 \mbox{ns}$ 
for the parameter values given in Table 
\ref{table_params_diffusion_compare}.  
From the results, we see that
the estimate given in equation \ref{equ_error_e_prt_large_time_step}
quantifies the error well for 
$\max 1/\alpha_{\mathbf{k}} \ll \Delta{t} \ll \tau_{\small mov}$,
where $\tau_{\small mov} \approx a^2/D = 1.95\times 10^{5} \mbox{ns}$.
}
\label{figure_nonrigorous_error_e_prt_large_time}
\end{figure}

In Figure \ref{figure_nonrigorous_error_e_prt_large_time}, the
theoretical error estimate 
\ref{equ_error_e_prt_large_time_step} for the elementary 
 particle
positions over a long time step is compared with the 
results of a numerical 
simulation in the case that there are no particle forces $(F^* = 0)$.
The numerical results were obtained from simulations of the 
fluid-particle system with physical parameters in Table 
\ref{table_params_diffusion_compare}.  In the comparison, 
the factors $Q$ were computed from the 
theoretical analysis with values given in Appendix 
\ref{appendix_constants_accuracy_error}.
The numerical simulations show good 
quantitative agreement with the formal error estimate 
\ref{equ_error_e_prt_large_time_step}.  We remark that 
the estimate is to be understood as 
an approximation and not a rigorous upper bound.
This agreement is evidence of the validity of the formal analysis.
As discussed in Section~\ref{section_error_estimate_small_time}, % PK
the errors arising in the presence of forces are not as explicitly quantifiable.
We leave further discussion and verification of the estimates
to future work in the context of specific applications.

\subsection{Error Estimates for Time Steps
which Underresolve Only Some Fluid Modes}
A key feature of the stochastic numerical scheme proposed in this 
work is that time steps can be chosen which only partially resolve 
the fluid dynamics.  That is, the method need neither resolve all 
of the velocity modes nor completely neglect the
inertia of the velocity field 
(as in Brownian/Stokesian dynamics~\citep{jfb:sd,as:asds,dle:bdhi,ts:mms}).
Rather, the time step can be chosen as needed to resolve the appropriate 
degrees of freedom 
of the fluid-particle system, having the fluid and thermal 
fluctuations interact appropriately with the structures.
The case is now 
discussed in which 
the time step $ \Delta t $ falls within the intermediate regime
\begin{equation}
\min\frac{1}{\alpha_{\mathbf{k}}} \lesssim \Delta{t} \lesssim 
\max\frac{1}{\alpha_{\mathbf{k}}},
\label{equ_time_step_int}
\end{equation}
where the dynamics of the fluid is only partially resolved.
It turns out that the 
error estimates for the fully resolved 
regime (Subsection~\ref{section_error_estimate_small_time}) and the underresolved 
regime (Subsection~\ref{section_error_estimate_large_time}) each serve separately 
as formal upper bounds for all time steps, including the intermediate 
regime~\ref{equ_time_step_int}.   Intuitively, then, one expects the numerical method 
to behave accurately over this intermediate range of time scales as well.  
To provide 
more quantitative support for this statement, error estimates are developed for 
the asymptotic regime
\begin{equation}
\frac{\rho a^{2}}{\mu} \ll \Delta{t} \ll \max \frac{1}{\alpha_{\mathbf{k}}}. 
\label{equ_time_step_int_asy}
\end{equation}

Then, for fluid modes that are well resolved, we have
\begin{eqnarray}
&& \\
\nonumber
\hat{e}_{\mbox{\small fld},\mathbf{k}}(\Delta{t})
& \approx & 
% (prefactor we discussed, double check this edit to make sure
% this is modified correctly)
\frac{M F^*}{\rho}
\delta_{a,\mathbf{k}}^*
\left(\frac{M}{\ell_F} + C\frac{1}{a}\right)
\left(
C' v_{\mbox{\small frc}} \Delta{t}^{2}
+ 
C'' \sqrt{D} \Delta{t}^{3/2}
\right) 
\mbox{ \hspace{0.15cm} if } \alpha_{\mathbf{k}} \Delta t \ll 1,
\end{eqnarray}
while the underresolved modes (with $ \alpha_{\mathbf{k}} \Delta t \gg 1 $) have 
the same error estimate~\ref{equ_error_e_fld_large_time_step} as in the fully 
underresolved case.

The errors incurred in the physical space variables describing the velocity and 
elementary particle positions can, in the asymptotic 
regime~\ref{equ_time_step_int_asy}, be estimated as
\begin{eqnarray}
&& \\
\nonumber
e_{\mbox{\small fld}}(\Delta{t})
& \approx & 
\frac{M  F^{*}}{\rho \nu^{3/4} L^{3/2}}
\left(
\frac{M}{\ell_F}
+ 
%\frac{\Delta{x}^3}{a^4}
C\frac{1}{a}
\right)
\left(
C'
v_{\mbox{\small frc}}\Delta{t}^{5/4}
+ 
C''
\sqrt{D}\Delta{t}^{{3}/{4}}
\right)
\end{eqnarray}

\begin{eqnarray}
&& \\
\nonumber
e_{\mbox{\small prt}}(\Delta{t}) 
& \approx &
Q_2
\frac{D}{a}
\Delta{t}
\\
\nonumber
& + &
\left(
\frac{M}{\ell_F}
+ 
C'
\frac{1}{a}
\right)
\left(
\sqrt{D}
v_{\mbox{\small frc}}
\Delta{t}^{{3}/{2}} + v_{\mbox{\small frc}}^{2} \Delta{t}^{2}\right),
\end{eqnarray}
where $ \nu = \mu/\rho $.

These errors can be compared
with the size of the actual changes in the system 
variables over a time step in the regime~\ref{equ_time_step_int_asy}, which can be
estimated by the same formulas as~\ref{equ_var_changes_large_time} except 
that the resolved velocity modes have changes of approximate size
\begin{eqnarray}
|\delta \hat{\mathbf{u}}_{\mathbf{k}}| & \approx & 
C \frac{M F^{*} \delta^{*}_{a,k}}{\rho} \Delta t + C' 
\frac{v_{\mbox{\small thm}}\sqrt{\alpha_{\mathbf{k}} \Delta t}}{N^{3/2}}.
\end{eqnarray}
The ratio of the errors to the corresponding magnitudes of the actual changes 
of the system variables over a time step is controlled by sums and products of
the nondimensional quantities 
$ \sqrt{D \Delta t}/{a} $, $ v_{\mbox{\small frc}} \Delta t/a $, 
$ (\nu \Delta t)^{1/4}/L^{1/2} $, 
$ M \frac{a}{L}$, 
and $ \frac{M a}{\ell_{F}} $.   
The former three remain small in the asymptotic
regime~\ref{equ_time_step_int_asy} under consideration, while the last two nondimensional 
groups (independent of time step) are related to our somewhat pessimistic bound on the force 
errors, as discussed in 
Subsection~\ref{section_error_estimate_small_time}.    The numerical 
method is thereby shown to remain theoretically accurate within this 
intermediate asymptotic regime.
In the absence of particle forces ($F^{*} = 0$), the error estimates become 
identical to those for the unresolved fluid regime 
(Subsection~\ref{section_error_estimate_large_time}).

One could also study the intermediate asymptotic regime
\begin{equation}
\min\frac{1}{\alpha_{\mathbf{k}}} \ll \Delta{t} \ll \frac{\rho a^{2}}{\mu},
\end{equation}
which exists only when $ \Delta x \ll a $.  For these time steps, all error 
estimates presented in Subsection~\ref{section_error_estimate_small_time} for 
the fully resolved regime remain valid, except that the estimate for the individual 
underresolved fluid modes is altered to
\begin{eqnarray}
\hat{e}_{\mbox{\small fld},\mathbf{k}}(\Delta{t})
& \approx & 
\frac{M F^* L^2}{\mu |\mathbf{k}|^2}
\left(
\frac{M}{\ell_F}
+ 
%\frac{\Delta{x}^3}{a^4}
C\frac{1}{a}
\right)
\delta_{a,\mathbf{k}}^*
\left(
C' v_{\mbox{\small frc}} 
+ 
C'' v_{\mbox{\small thm}} 
\right)
\Delta{t}. 
\end{eqnarray}
As with the other regimes, the errors in this regime are small relative to the magnitude 
of the changes of the actual system variables over a time step.

By simple extension of the above arguments for time steps falling at the transitions 
between the asymptotic regimes, we see that 
the numerical method has been designed to remain theoretically 
accurate for all time steps $ \Delta t \ll \tau_{\mbox{\small mov}} (a) $, 
regardless of how well the fluid dynamics are resolved.  

% so the numerical comparison 
%for the intermediate time step regime is plotted along with the large time step 
%regime in Figure~\ref{figure_nonrigorous_error_e_prt_large_time}.

\section{Physical Behavior of the Method and Numerical Results}
\label{section_numerical_results}
To ensure that the immersed boundary method with
thermal fluctuations serves as a plausible 
physical framework for modeling microscale systems, we
verify that the method exhibits several fundamental 
features which are correct according to the laws of 
statistical physics \citep{IBler:mcsp}.
In Subsection \ref{section_diffusion_coeff}
an expression for the diffusion coefficient of
immersed particles is derived, and it is shown
that in three dimensions the mean squared
displacement scales linearly in time 
and inversely in the particle size.
It is
further shown in Subsection \ref{section_V_decay}
 that
the stochastic immersed boundary method captures
the correct $\tau^{-{3/2}}$ 
power law for the decay of the tail of the autocorrelation 
function of the particle velocity~\citep{yp:tdcfm,bja:dvaf,
hinch1975, dorfman1970,ernst1971,hauge1973,mazur1974,cohen1974}.

In determining the thermal forcing in Subsection 
\ref{section_thermal_forcing}, 
we imposed the requirement that the degrees of freedom of the fluid obey 
Boltzmann statistics in thermal equilibrium.  In fact, the complete 
system including immersed structures 
should obey Boltzmann statistics.  
In Subsection 
\ref{section_equil_statistics}, we study
the equilibrium statistics of 
immersed particles subject to a conservative
force and show through numerical simulation that 
they do exhibit the correct Boltzmann statistics.  
To demonstrate some applications and as a further 
verification of the physical plausibility of the method,
it is shown in Subsections \ref{section_osmoticNoninteracting} -- 
\ref{section_application_knots}
how the method can be used to model osmotic effects such as 
the pressure of confined particles, dimers, and 
polymers~\citep{go:doff, einstein1956}. 
In Subsection \ref{section_application_motor},
another application to a basic model 
of a molecular motor protein immersed in a fluid 
subjected to a hydrodynamic load force is presented 
~\citep{peskin1993}.

\subsection{Diffusion of Immersed Particles}
\label{section_diffusion_coeff}
In this section the diffusion of particles in
the stochastic immersed boundary method is discussed
and an expression for the diffusion coefficient is derived.  
As part of the analysis it is shown that the correct 
diffusive scaling is obtained for three dimensional systems.
%It is further shown that for the tail of the autocorrelation
%function for the particle velocity, the $\tau^{-{3/2}}$ power 
%law is captured.  
To verify the validity of the 
approximations made in the analysis and to demonstrate the 
applicability of these results in practice, the 
results of the analysis are compared to the results of 
numerical simulations. 

In three dimensions, 
 the diffusion coefficient for a single particle 
is defined as 
\begin{eqnarray}
\label{equ_diffusion_est1}
D & = & \lim_{t \rightarrow \infty}
 \frac{\left<| \mathbf{X}(t) - \mathbf{X}(0) |^2\right>}{6t}. 
\end{eqnarray}
In the notation the superscripts on the particle position are
suppressed since only a single immersed particle is considered.

An estimate for the diffusion coefficient of a 
single particle (with no interactions with 
other particles) in the stochastic immersed boundary 
method is derived in Subsection \ref{section_deriv_diff_coeff}
from the autocorrelation function of the velocity field of
the fluid.  This estimate can be expressed as
\begin{eqnarray}
\label{equ_diffusion_est3}
D & = & \frac{k_B{T} L^3}{3\rho}\sum_{\mathbf{k}} 
\frac{|\hat{\delta}_{a,\mathbf{k}}|^2\Upsilon_{\mathbf{k}}}
{\alpha_{\mathbf{k}}},
\end{eqnarray}
where $\Upsilon_{\mathbf{k}}$ is defined in appendix 
\ref{appendix_velocity_autocorrelation} 
and
$\hat{\delta}_{a,\mathbf{k}}$ is defined in
appendix \ref{appendix_delta_func_hat}.  This diffusivity 
as simulated by the stochastic immersed boundary method 
exhibits the physically correct scaling with respect to 
physical parameters \citep{kramer2003}.

The diffusion coefficient is estimated from the numerical simulations using
\begin{eqnarray}
\label{equ_diffusion_num_est}
\tilde{D} & \approx & \frac{1}{6n t_1} 
\sum_{m = 1}^{n} \left |\mathbf{\tilde{X}}^{m}(t_1) 
- \mathbf{\tilde{X}}^{m}(0) \right |^2,
\end{eqnarray}
where $n$ is the number of sampled trajectories 
of fixed duration $t_1$.
The notation $\mathbf{\tilde{X}}^{m}$ denotes 
the simulated particle position from the $m^{th}$ trajectory.

\begin{figure}[h*]
\centering
\epsfxsize = 4in
\epsffile[0 0 360 290]{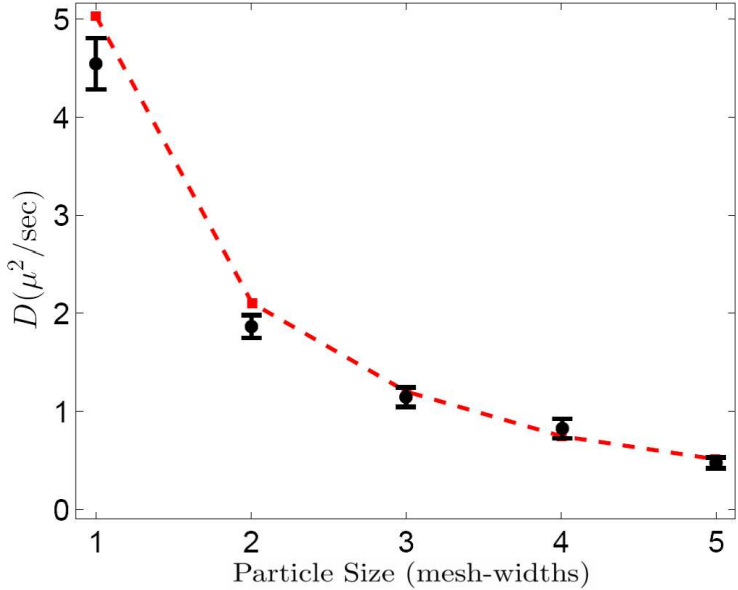} %PJA: changed figure to be microns^2/sec
\caption[Comparison of Diffusion Coefficients of IB Particles, Theory vs. 
Numerical Simulation]
{A comparison of the analytic estimate \ref{equ_diffusion_est3}
for particles of the IB method with numerically estimated diffusion
coefficients for the parameters given in Table 
\ref{table_params_diffusion_compare}.  The dashed line indicates the analytic
estimate \ref{equ_diffusion_est3}.  The data points denote estimates of the 
diffusion coefficient obtained from numerical simulations using the method.
The error bars
denote one standard deviation of the sampled values.
In the numerical simulations long time steps satisfying
$\max 1/\alpha_{\mathbf{k}} \ll \Delta{t}$ were taken
which under
For the parameters values given in
Table \ref{table_params_diffusion_compare}
the time step $\Delta{t} = 10^3 \mbox{ns}$
was used in the numerical simulations, where
$\max 1/\alpha_{\mathbf{k}} = 25.4\mbox{ns}$.  The
particle size $k_0$ corresponds to the parameter value 
$a = k_0\Delta{x}$ of the particle representation function 
$\delta_a$ defined in appendix \ref{appendix_delta_func}, 
where $k_0$ controls the number of mesh-widths spanned by 
the $\delta_a$ function.} %PJA: changed figure caption (last line)
\label{figure_diffusion_compare1}
\end{figure}

In Figure \ref{figure_diffusion_compare1}, the theoretical 
estimate of the diffusion coefficient as given in equation
\ref{equ_diffusion_est3} is compared to the 
numerical estimate given in equation \ref{equ_diffusion_num_est} 
for particles with sizes $a = 1, 2, 3, 4, 5$ and for a long
time step which underresolves the dynamics of the fluid.
For the parameters of the fluid-particle system used in the 
numerical simulations, see table 
\ref{table_params_diffusion_compare}.  For each particle
size, the numerical estimates were made from $n = 10^4$ 
sampled trajectories with $\Delta{t} = 10^3 \mbox{ns}$ and 
$t_1 = 10^{4} \mbox{ns}$.
We remark that in the simulations,
while the time step underresolves the fastest modes of the
fluid, there is still good agreement between the diffusion
coefficient of the simulated particle and the
theoretical estimate
\ref{equ_diffusion_est3}.
A tabulation of
the theoretical diffusion coefficient as given in 
equation \ref{equ_diffusion_est3} for the 
immersed boundary method with various ratios of $ a/\Delta x $ and
$ a/L $ can be found in \citep{kramer2003}.

\subsubsection{Derivation of the Diffusion Coefficient} 
\label{section_deriv_diff_coeff}
To derive an analytic estimate for the diffusion coefficient
the semidiscretized equations \ref{semidiscretization_fluid_equ1}
and \ref{semidiscretization_particle_equ0} are considered.  
Alternatively, the diffusion coefficient can also be derived 
directly from the stochastic immersed boundary equations by
a stochastic mode reduction procedure~\citep{prk:smrib}.
In equations \ref{semidiscretization_fluid_equ1} and 
\ref{semidiscretization_particle_equ0} the 
immersed particle dynamics are given by
\begin{eqnarray}
\frac{d\mathbf{X}(t)}{dt} & = & \mathbf{U}(\mathbf{X}(t),t), 
\end{eqnarray}
where $\mathbf{U}$ is given in Fourier space by
\begin{eqnarray}
\mathbf{U}
(\mathbf{x},t) 
	            & = & \sum_{\mathbf{k}} L^3 \hat{\delta}_{a,\mathbf{k}}(\mathbf{x})
                          \hat{\mathbf{u}}_{\mathbf{k}}(t) 
                         \exp\left(
                         i2\pi{\mathbf{x}}\cdot\mathbf{k}/L
                         \right),
\end{eqnarray}
with the coefficient
$\hat{\delta}_{a,\mathbf{k}}(\mathbf{x})$ defined
in appendix \ref{appendix_delta_func_hat}.

The autocorrelation function of the velocity of an immersed particle is 
\begin{eqnarray}
R(t,t + \tau) & = & \left<\mathbf{U}\left(\mathbf{X}(t)
,t\right)\cdot
\mathbf{U}\left(\mathbf{X}\left(
t + \tau
\right),t + \tau \right)\right>. 
\end{eqnarray}

To estimate this, we use the fact that the statistics 
of $ \mathbf{U}(\mathbf{x},t) $ are approximately shift 
invariant in both $t$ and $\mathbf{x}$, with the 
approximation improving as the spatial grid is refined.  
We further assume that the time scale associated with the 
fluid velocity is small relative to the time scale of 
the immersed particle motion, so that a particle moves 
a negligible distance relative to its size $a$ 
during the correlation time of the fluid 
velocity.  These approximations give 
\begin{eqnarray}
R(t,t + \tau) & \approx & \left<\mathbf{U}(0,0)\cdot
                          \mathbf{U}(0,\tau)\right> := R(\tau). 
\end{eqnarray}

By applying (\ref{equ_autocorrolation_1}), the autocorrelation function of the velocity of an immersed 
particle can then be expressed
as
\begin{eqnarray}
\label{equ_diffusion_auto_corr_deriv}
R(\tau)      
%\nonumber
              & \approx & \sum_{\mathbf{k},\mathbf{k}'}L^6 
                    \overline{\hat{\delta}}_{a,\mathbf{k}} 
                    \hat{\delta}_{a,\mathbf{k}'}
                    \left<\overline{\mathbf{u}_{\mathbf{k}}}(t) 
                              \cdot \mathbf{u}_{\mathbf{k'}}(t + \tau) 
                    \right> \\
\nonumber
              & = & 
\sum_{\mathbf{k}\in K}
                      L^6 |\hat{\delta}_{a,\mathbf{k}}|^2 
                      3
                      \frac{D_k}{\alpha_{\mathbf{k}}}
                     e^{-\alpha_{\mathbf{k}} |\tau|}  \\
\nonumber
& + & 
\sum_{\mathbf{k}\not\in K}
                      L^6 |\hat{\delta}_{a,\mathbf{k}}|^2 
                      4
                      \frac{D_k}{\alpha_{\mathbf{k}}}
                      e^{-\alpha_{\mathbf{k}} |\tau|} \\
\nonumber
              & = & 
            \frac{k_B{T} L^3}{\rho}
\sum_{\mathbf{k}} 
|\hat{\delta}_{a,\mathbf{k}}|^2 
\Upsilon_{\mathbf{k}}
e^{-\alpha_{\mathbf{k}}|\tau|}, 
\end{eqnarray}      
where $\Upsilon_{\mathbf{k}}$ is defined in appendix 
\ref{appendix_velocity_autocorrelation}.

An important point for the numerical method developed 
in Section 
\ref{section_numerical_method}
is that this structure of the correlation
function is preserved even for finite time steps, provided 
only that the time step is small enough that the immersed elementary 
particles do not move significantly during a time step 
($ \Delta t \ll \tau_{\mathrm{diff}} (a)$), where
$\tau_{\mbox{\tiny diff}}(a)$ is the
time scale of a particle to diffuse over a distance 
equal to its size $a$.  Were we to have 
used instead a numerical method based on a stochastic Taylor 
expansion \citep{platen1992},
we would have to restrict the time 
step $ \Delta t $ to be small enough so that $ R(\tau) $ is well 
approximated by a Taylor expansion for 
$ |\tau | \lesssim \Delta t $, 
which would add the additional restriction that 
$ \Delta t \ll 1/\alpha_{\mathbf{k}} $.  
Our more accurate representation for the fluid dynamics over 
a time step allows us to obviate this other condition, as 
demonstrated in Figure \ref{figure_diffusion_compare1}.  
% The first equality follows similarly to the derivation
%in appendix \ref{appendix_velocity_autocorrelation},
%where it is used that for
%two distinct modes $\mathbf{k}$ and $\mathbf{k'}$ which are 
%conjugate in \ref{equ_DFT_real_constr} the expectation of the
%product is zero.
%%$<\bar{\mathbf{u}}_{\mathbf{k}}(s)\cdot
%%\mathbf{u}_{\mathbf{k'}}(r)> = 0$.  
%The second equality follows by substitution of 
%the autocorrelation function of the velocity field of the fluid, 
%derived in \ref{equ_corr_deriv_2}, and by
%carefully taking into account the constraints 
%\ref{equ_DFT_incompressible}
%and 
%\ref{equ_DFT_real_constr}.

A useful identity relating the autocorrelation function to
the mean squared displacement of an immersed particle is
\begin{eqnarray}
\left< | \mathbf{X}(t) - \mathbf{X}(0) |^2\right> & = & \left<
 \int_0^{t} \frac{d\mathbf{X}(s)}{ds} ds \cdot
 \int_0^{t} \frac{d\mathbf{X}(r)}{dr} dr  \right> \\ 
 \nonumber
                 & = & 2 \int_0^{t} R(r)\cdot(t - r) dr. 
\end{eqnarray}

This allows for the diffusion coefficient to be estimated by the 
Kubo formula~\citep{IBrk:sp2}: 
\begin{eqnarray}
\label{equ_diffusion_est2}
D & = & \frac{1}{3}\int_0^{\infty} R(r) dr,
\end{eqnarray}
By substituting the estimate \ref{equ_diffusion_auto_corr_deriv} 
into \ref{equ_diffusion_est2} and evaluating the integral,  
the expression \ref{equ_diffusion_est3} for the 
diffusion coefficient is obtained.

\subsection{Algebraic Decay of Velocity Autocorrelation Function}
\label{section_V_decay}
For immersed particles diffusing in a viscous fluid, 
 the particle
motion is strongly coupled to the motion of the fluid.  
As a particle moves along a particular direction, fluid is dragged along 
with it.  When the particle changes direction, it is
resisted by a viscous force arising from its motion relative to the 
nearby fluid with momentum related to the recent past of the 
particle's motion.  This induces a somewhat stronger memory in the 
particle velocity than a standard model based on a constant Stokes 
drag would predict.
In particular, a careful analysis of physical 
Brownian motion,
including a more detailed model for 
the force between an immersed particle and the surrounding fluid, 
yields for $ D \ll \mu/\rho $
~\citep{sorensen2005,yp:tdcfm,bja:dvaf,
hinch1975, dorfman1970,ernst1971,hauge1973,mazur1974,cohen1974}
\begin{equation}
\label{equ_alg_decay_R_theory}
R(\tau) \approx \frac{k_{B} T \rho^{1/2}}{4\mu^{3/2}} \tau^{-3/2} \mbox{ for }
\tau \gg \rho a^{2}/\mu.
\end{equation}
The condition $ D \ll \mu/\rho $ can 
readily be 
checked to hold for typical microbiological systems.

For the stochastic immersed boundary method, 
it is shown in Subsection \ref{section_deriv_algebraic_decay}
that this general behavior is recovered with   
\begin{equation}
%\label{equ_alg_decay_R_ib}
R(\tau)\approx 
\left[C_{\mbox{\tiny IB}}
\frac{k_B{T} \rho^{1/2}}{\mu^{3/2}}
\right]
\tau^{-{3}/{2}} \mbox{ for } \rho a^{2}/\mu \ll \tau \ll \rho L^{2}/\mu, 
\label{equ_tail}
\end{equation}
where $C_{\mbox{\tiny IB}} = \frac{1}{4 \pi^{{3}/{2}}}$. 
The constant prefactors differ slightly due to
the different ways particles are represented in the 
physical model and immersed 
boundary method.

The restriction that 
$ \tau \ll \rho L^{2}/\mu $ for the $ \tau^{-3/2} $ scaling 
in the immersed boundary method is a finite 
size effect which should have an analogue for physical Brownian motion.
For very long times where $ \tau \gg \rho L^{2}/\mu $
the correlation function $ R(\tau) $  decays
exponentially, with rate governed by that of the lowest 
wavenumber modes in the Fourier series 
(\ref{equ_diffusion_auto_corr_deriv}).  However, by these times
the autocorrelation function would already be very small so this
very long time regime is of little practical interest.

These results show that the decay of the particle velocity 
autocorrelation function 
in the stochastic immersed boundary method has the correct scaling with 
respect to time and physical parameters.

\subsubsection{Derivation of Algebraic 
Decay of Velocity Autocorrelation Function}

\label{section_deriv_algebraic_decay}
In this discussion, the reference to wavenumbers $ \mathbf{k} $  
implicitly indicates the value within 
the equivalence class of aliased wavenumbers such that each 
component $ |\mathbf{k}^{(j)}| \leq N/2 $.  For this purpose 
one can choose any scheme to select a unique value 
when $ \mathbf{k} $ lies on the boundary of this set.

First observe from the scaling properties of Fourier transforms 
and the definition of $ \delta_{a} $ from (\ref{equ_phi_def_delta_a}) that 
\begin{equation}
 \hat{\delta}_{a,\mathbf{k}} \approx \hat{\delta}_{a,\mathbf{0}} 
= \frac{1}{L^{3}} \mbox{ for } |\mathbf{k}| \ll L/a, 
\end{equation}
and $ \hat{\delta}_{a,\mathbf{k}} $ decays rapidly with respect 
to $ |\mathbf{k}|a/L $.  Along with the fact that the high wavenumber components
of $ R(\tau) $ 
decay at a faster rate $ \alpha_{\mathbf{k}} $ than the low wavenumber components, 
it then follows from the Fourier series 
representation~\ref{equ_diffusion_auto_corr_deriv} 
for the particle velocity autocorrelation function that the sum will be
dominated by the terms with $ |\mathbf{k}| \lesssim L/a $

Over the intermediate asymptotic time interval indicated 
in~\ref{equ_tail}, the 
time $ t $ is small compared to the decay time 
$ 1/\alpha_{\mathbf{k}} \sim  \rho L^{2}/\mu$ of the low 
wavenumber modes $ |\mathbf{k}| \sim 1 $, but large compared to 
the decay time $ 1/\alpha_{\mathbf{k}} \sim \rho a^{2}/\mu $ of 
the (relatively high) wavenumber modes $ |\mathbf{k}| \sim L/a $ 
corresponding to the length scale of the particle.   Combining these 
observations, 
there exists a time-dependent wavenumber scale $ k_{c} (t) $ 
which satisfies $ 1 \ll k_{c} (t) \ll N/2 $ such that 
$ e^{-\alpha_{\mathbf{k}} t} \approx 1 $ for $ |\mathbf{k}| \ll k_{c} (t) $ 
and $ e^{-\alpha_{\mathbf{k}} (t)} \approx 0 $ for wavenumbers such that 
$ |\mathbf{k}| \gg k_{c} (t) $.  Consequently, over the intermediate 
asymptotic 
time interval, the Fourier series~\ref{equ_diffusion_auto_corr_deriv} 
is dominated
 by contributions from wavenumbers 
$ 1 \leq |\mathbf{k}| \lesssim k_{c} (t) \ll N/2 $.  
These observations allow us to make the following simplifying 
approximations over the 
time interval $ \rho a^{2}/\mu \ll t \ll \rho L^{2}/\mu $:
\begin{itemize}
\item The prefactors multiplying the exponential in each Fourier 
series term 
may be approximated by their low wavenumber limits:
 \begin{equation}
 |\hat{\delta}_{a,\mathbf{k}}|^{2} \Upsilon_{\mathbf{k}} \approx 2/L^{6}   
\mbox{ for }
  |\mathbf{k}| \ll N/2. 
\end{equation}
\item The decay rate in the exponential may be approximated
for $ |\mathbf{k}| \ll N/2 $ by its low
wavenumber asymptotics
\begin{equation}
\label{equ_alg_replace}
 \alpha_{\mathbf{k}} \approx A |\mathbf{k}|^{2}; A =
  4 \pi^{2} \mu \rho^{-1} L^{-2}.
\end{equation}
 \item The Fourier sum may be extended to the full integer lattice, 
because with the replacement \ref{equ_alg_replace}, the additional 
terms for large wavenumbers will be exponentially small 
and make 
a negligible contribution.
 \item This Fourier sum over the integer lattice can be approximated 
by an integral over continuous $ \mathbf{k} $, because the dominant 
contribution comes from a large number of lattice sites 
$ 1 \leq |\mathbf{k}| \lesssim k_{c} (t) $, with $ k_{c} (t) \gg 1 $.  
 \end{itemize}
 
Applying these simplifications and then changing to spherical 
coordinates with radial variable $ k = | \mathbf{k}| $, we obtain 
\begin{eqnarray}
R(\tau) & \approx & \int_{\mathbb{R}^{3}} \frac{2 k_{B} T }{\rho L^{3}} 
\exp (-  A  |\mathbf{k}|^{2} \tau) \, d \mathbf{k} \\
\nonumber
& = & \frac{8 \pi k_{B} T }{\rho L^{3}} \int_{0}^{\infty} k^{2} 
\exp (-  A  k^{2} \tau) \, dk \\
\nonumber
& = & \frac{8 \pi k_{B} T }{\rho L^{3}} \frac{1}{2}
\left(\sqrt{2\pi \frac{1}{2A\tau}}\frac{1}{2A\tau}\right), 
\end{eqnarray}
where the second equality follows readily by
using standard facts about Gaussians.  In particular,  
the integral
can be treated as the expectation of the second moment by
introducing the standard normalization factor.
Using \ref{equ_alg_replace} 
and simplifying the 
expression yields \ref{equ_tail}.

\subsection{Equilibrium Statistics of Immersed Particles}
\label{section_equil_statistics}
For the particle-fluid system with the fixed temperature 
$T$, volume $V$, and number of elementary 
particles $M$, with the particles subject to a 
conservative force field, we have from statistical 
mechanics that the equilibrium probability density $ \Psi $ of 
the elementary 
 particle positions should have Boltzmann statistics:
\begin{eqnarray}
\Psi(\{\mathbf{X}\}) 
& = & \frac{1}{Z} 
\exp\left(-\frac{\mathcal{E}(\{\mathbf{X}\})}
{k_B{T}}\right), 
\end{eqnarray}
where $\mathcal{E}$ is the energy of a configuration of 
elementary 
 particles.
The factor $Z$ is the normalization factor so that the density 
integrates to one.  The Boltzmann distribution arises 
from the thermodynamic condition that the equilibrium 
probability distribution of the microscopic states 
maximize 
%Gibbs 
entropy
while maintaining a 
fixed average energy for the system
\citep{IBler:mcsp}.

\begin{figure}[h*]
\centering
\epsfxsize = 4in
\epsffile[69 212 506 655]{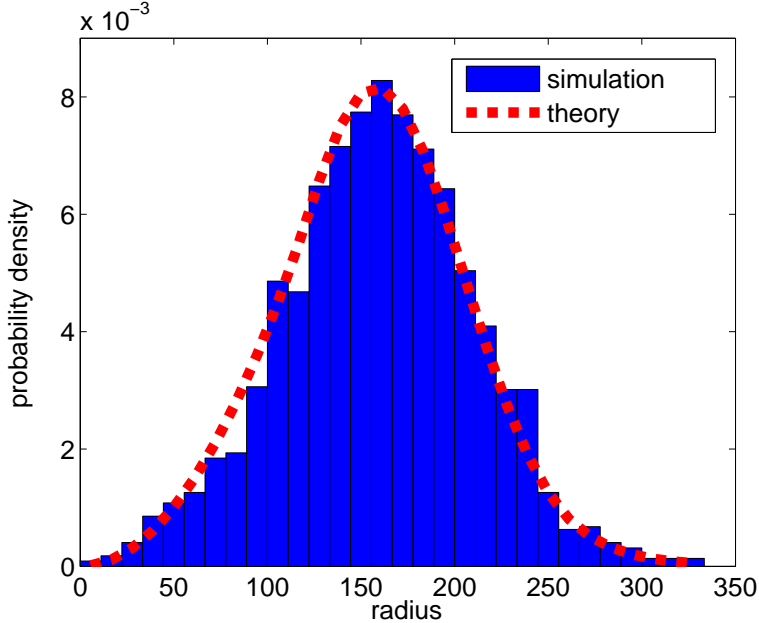}
\caption[Boltzmann Distribution vs Numerical Equilibrium Distribution]
{Boltzmann Distribution vs Numerical Equilibrium Distribution.  
The spherical potential energy has the parameters
$R_1 = 125$, $R_2 = 250$, $c = 6 {k_B{T}}/(R_2 - R_1)$.  The 
simulation was run with the parameters
   $N = 16 $, 
   $L = 1000 \mbox{ nm}$, 
   $\Delta x = L/N$, 
$\mu  = 6.02 \times 10^5 \mbox{ amu/(nm $\cdot$ ns)}$, 
$\rho = 602~\mbox{amu/$\mbox{nm}^3$}$, $T = 300 \mbox{ K}$,
$\Delta{t} = 1000 \mbox{ ns}$. 
$200,000$ time steps were simulated.}
\label{figure_sphere1_boltzmann_vs_numerical}
\end{figure}

In deriving the thermal forcing of the system
in Subsection \ref{section_thermal_forcing}, only the energy 
associated with the fluid modes was considered.  
When immersed structures are subject to a force,
it is not entirely clear that the correct equilibrium 
distribution for the system as a whole will be attained.

When formulated in continuous time,
the fluid-particle coupling in the
immersed boundary method conserves energy 
exactly~\citep{peskin2002}.  
While 
this may suggest that the correct equilibrium statistics should be 
obtained (up to the appropriate definition of an effective 
temperature), the discretization of time in the numerical 
method could in principle disrupt it, particularly since we 
are not employing a symplectic 
method~\citep{ss:nhp,gnm:nmssp}.

We now show that  the numerical method from 
Section~\ref{section_summary_num_method} appears 
to yield results consistent with the
Boltzmann distribution, at least for the statistics of 
the position of a single immersed particle.  For a more rigorous 
approach in which the Fokker-Plank equations associated 
with the stochastic immersed boundary method are 
analyzed, 
 see the related work~\citep{atzberger2005b}.

To facilitate calculation of the equilibrium statistics, 
the particles are subject to a radially symmetric external force
depending on $r = \left|\mathbf{X}\right|$ with the potential 
energy
\begin{eqnarray}
\label{equ_spherical_energy}
V(r) & = & \left\{ \begin{array}{ll} 
                          0,                & r < R_1  \\
                          c\cdot (r - R_1), & R_1 \leq r \leq R_2 \\
                          c\cdot(R_2 - R_1),   & r > R_2. 
                         \end{array}
                 \right.
\end{eqnarray}
It is assumed that $c > 0$ so that this can be thought of physically 
as the potential associated with confining particles to a spherical chamber 
of radius $R_2$.  The inner radius $R_1$ is used to soften the
particle-wall interactions to avoid issues of numerical stiffness
and $R_2$ is taken significantly smaller than the spatial period 
of the lattice $L$.

For a single immersed particle,
 the Boltzmann distribution for its  radial coordinate is
\begin{eqnarray}
\label{equ_Boltzmannn_particle}
\tilde{\Psi}(r) 
& = & \frac{4\pi r^2}{\tilde{Z}} \exp\left(-\frac{V(r)}{k_B{T}}\right), 
\end{eqnarray}
where $\tilde{Z}$ is the normalization factor.  

In Figure \ref{figure_sphere1_boltzmann_vs_numerical}, the 
equilibrium statistics of immersed particles simulated with 
the numerical method are compared with the Boltzmann distribution
\ref{equ_Boltzmannn_particle}.  The simulations were
performed with 
$R_1 = 125 \mbox{nm}$, $R_2 = 250 \mbox{nm}$ and $c = 6k_B{T}/(R_2 - R_1)$ 
with the parameters of the fluid-particle system given in Table 
\ref{table_params_diffusion_compare}.

\subsection{Osmotic Pressure of Confined Non-interacting Particles}
\label{section_application_osmotic}
\label{section_osmoticNoninteracting}
Osmosis is a phenomenon that occurs in many microscale biological
systems.  When diffusing particles are confined to a chamber 
by a boundary which is permeable to fluid 
but less permeable to particles,
 a pressure difference develops 
between the  inside and the outside of the 
chamber.  This difference is referred to as the 
``osmotic pressure''.

When the confining boundary is impermeable to particles
and the system is in equilibrium, van't Hoff's law \citep{IBler:mcsp}
relates the osmotic pressure to the concentration of 
the confined particles as
\begin{eqnarray}
p_{\mbox{\small osmosis}} & = & \bar{c}_0 k_B T, 
\end{eqnarray}
where $\bar{c}_0$ is the number of particles per unit volume in
the chamber.  
More precisely, when the number of confined particles is small 
enough that the instantaneous pressure fluctuates, then 
 van't Hoff's law should describe the ensemble or time average of
the pressure difference that arises from confinement.

One should see a signature of
van't Hoff's law in the fluid pressure
when a collection of $ M$ non-interacting particles in a conservative force field with potential $ V $ 
are simulated
by  the stochastic immersed boundary method, given that the
method was shown in Subection~\ref{section_equil_statistics}
 to produce correct
Boltzmann equilibrium statistics.  Indeed, taking 
 the
expectation of the velocity, force, and pressure 
with respect to Boltzmann's distribution (ensemble average)
in the fluid equation 
\ref{equ_cont_fluid_equ} 
gives 
\begin{eqnarray}
\label{equ_pressure_1}
0 & = & -\nabla \langle p(\mathbf{x})\rangle
\mbox{ }  + \langle\mathbf{f}_{\mbox{\small prt}} 
(\mathbf{x})\rangle.
\end{eqnarray}
% PRK:  I fixed  a mistake with the M and moved the averaging bracket for p

This is obtained using that
$\langle \mathbf{u}\rangle = 0$ and 
$\langle \mathbf{f}_{\mbox{\small thm}}\rangle = 0$.
The notation $\langle \cdot\rangle$ denotes the ensemble 
average over the thermal fluctuations  
and $\langle p \rangle$ denotes 
the average of the fluid  %PRK
pressure field.  

The ensemble average of the conservative force field with potential
$V$ at location $\mathbf{x}$ is
\begin{eqnarray}
\label{equ_F_prt_osmosis}
\langle\mathbf{f}_{\mbox{\small prt}}(\mathbf{x})\rangle
& = & \frac{-M \nabla{V}(\mathbf{x})}{Z} 
e^{-\frac{V(\mathbf{x})}{k_B{T}}}. 
\end{eqnarray}
% PRK:  Fixed another M mistake

\begin{figure}[h*]
\centering
\epsfxsize = 4in
\epsffile[69 212 506 655]{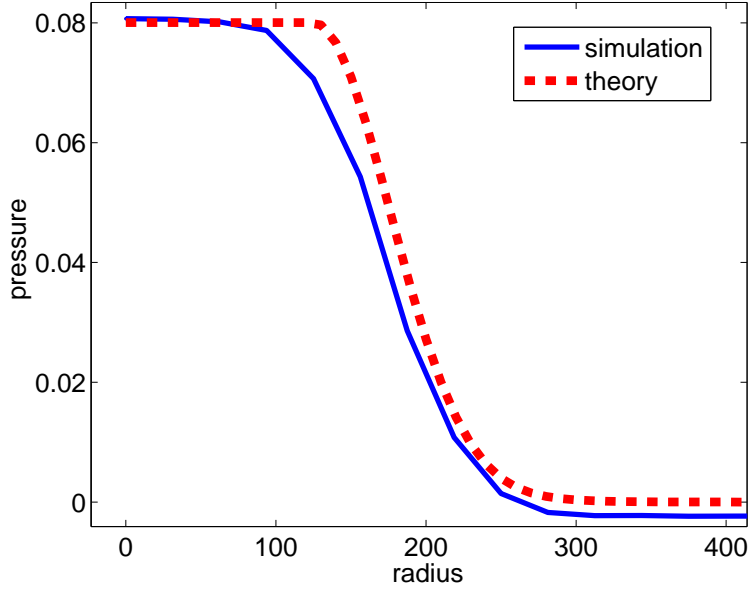}
\caption[Time Averaged Radial Pressure]
{Time Averaged Radial Pressure.
The spherical potential energy has the parameters
$R_1 = 125$, $R_2 = 250$, $c = 6 {k_B{T}}/(R_2 - R_1)$. 
 The simulation was run with the parameters
   $N = 16 $, 
   $L = 1000 \mbox{ nm}$, 
   $\Delta x = L/N$, 
$\mu  = 6.02 \times 10^5 \mbox{ amu/(nm $\cdot$ ns)}$, 
$\rho = 602~\mbox{ amu/$\mbox{nm}^3$}$, $T = 300 \mbox{ K}$,
$\Delta{t} = 1000 \mbox{ ns}$. 
$200,000$ time steps were simulated. 
}
%?edit?NOTE: difference may be explained by two sources
%delta function averages the force applied to the fluid.  
%The pressure was averaged radially over a discrete grid
%which may smooth data depending on the radial bin 
%into which the data falls}
\label{figure_sphere1_time_avg_radial_pressure}
\end{figure}

The average pressure $\langle p(\mathbf{x}) \rangle$ can 
be determined up to an additive
constant from \ref{equ_pressure_1} and \ref{equ_F_prt_osmosis}
by taking the line integral  
in $\mathbf{x}$.  From the fundamental theorem of 
line integrals with the additive constant set to zero 
(to give the appropriate decay 
at large $ \mathbf{x}$) 
we have
\begin{eqnarray}
\label{equ_osmosis_pressure}
\langle
p(\mathbf{x})
\rangle & = & \frac{M} 
{Z} e^{-\frac{V(\mathbf{x})}{k_B{T}}} k_B{T}
\end{eqnarray}
By the assumption of Boltzmann statistics for the immersed 
particles the concentration field is
\begin{eqnarray} 
c_0(\mathbf{x}) = \frac{M} 
{Z} e^{-\frac{V(\mathbf{x})}{k_B{T}}}. 
\end{eqnarray}
Substitution into \ref{equ_osmosis_pressure} and integrating over the 
chamber gives van't Hoff's law.  
This derivation also indicates that the stochastic immersed 
boundary method has a fluid 
pressure field which should respect 
the local formulation of van't Hoff's law:  
\begin{eqnarray}
p_{0}(\mathbf{x}) & = & c_0(\mathbf{x}) k_B T.
\end{eqnarray}

From a statistical mechanical point of view of this derivation,
this fluid pressure
can be thought of as arising 
from the fluctuations of the system which persistently subject a 
particle to the confining forces in the vicinity of the wall 
with a frequency determined by the Boltzmann statistics.  The 
forces are then transferred to the fluid by the viscous 
particle-fluid interactions.  Similar ``mesoscopic'' points 
of view of osmosis have
been used 
in~\citep{go:doff, einstein1956, atzberger2005c}.

In Figure \ref{figure_sphere1_time_avg_radial_pressure}, a comparison 
is made of the pressure predicted by a local van't Hoff's law 
taking into account the local solute concentration and 
the time average of the pressure field of the fluid obtained in a
numerical simulation using the stochastic immersed boundary method for
an immersed particle confined by the spherically symmetric potential 
given by equation \ref{equ_spherical_energy}. 

Another quantification of osmotic pressure is the average force per unit area 
exerted on the walls of the chamber which confine the solute.  For a 
spherical chamber $ \Omega $ 
of radius $R$ in which the solute exerts a (generally repulsive) areal 
force density $ \mathbf{F}_{\mathrm{pw}} (\mathbf{z}) $ on a portion of 
the wall boundary $ \partial \Omega$  at relative location $ \mathbf{z} $, 
this osmotic pressure is given by: 
\begin{eqnarray}
\label{osmosis_p_wall_equ}
p_{\mbox{wall}} & = & 
\frac{1}{4\pi R^2}
\int_{\Omega}
\int_{\partial \Omega} 
\mathbf{F}_{\mathrm{pw}}(\mathbf{y} - \mathbf{x}) \cdot \frac{\mathbf{y}}{|\mathbf{y}|} c (\mathbf{x}) \, 
d\mathbf{y} \, d \mathbf{x}
\end{eqnarray}
where $ c(\mathbf{x}) $ denotes the average concentration of the solute particles.  Note that this formula applies also when the solute particles interact with each other.  We will only be considering isotropic wall-solute interactions (and uniform distribution of wall molecules) so that we can write the force density  in terms of a (typically nonnegative) scalar function $ q_{\mathrm{pw}} (r) $:
 $ \mathbf{F}_{\mathrm{pw}} (\mathbf{z})
 = q_{\mathrm{pw}}(|\mathbf{z}|) \mathbf{z}/|\mathbf{z}| $ and the concentration density as $ c(\mathbf{x}) = c (|\mathbf{x}|)$. 
 The expression in equation \ref{osmosis_p_wall_equ} for the osmotic pressure  on the chamber wall 
 %PRK:  word order
 can then be simplified by integrating over the angular degrees of freedom: 
\begin{equation}
p_{\mbox{wall}}  = 
\frac{1}{R^2}
\int_0^R
h(r)
c(r)
r^2
dr  
\label{p_wall_equ}
\end{equation}
with
\begin{eqnarray}
\nonumber
h(|\mathbf{x}|) & = & \int_{\partial \Omega}
\mathbf{F}_{\mathrm{pw}}(|\mathbf{y} - \mathbf{x}|)\cdot \frac{\mathbf{y}}{|\mathbf{y}|}
d\mathbf{y} \\
& = & \frac{\pi}{r} \int_{R - r}^{R + r} q_{\mathrm{pw}}(\rho) \left(\rho^2 + R^2 -r^{2}\right) d\rho. 
\label{def_h_r}
\end{eqnarray} 
The function $h(r) $ 
can be interpreted as the integrated normal force applied to the wall
by a solute particle located a distance $ r = |\mathbf{x}| $ from the origin.  The change 
of variable used to obtain the last integral was $\rho = |\mathbf{y} - \mathbf{x}|$.

When the potential confining the solute is ``hard-walled,'' 
in the
sense that the solute-wall interactions occur only in a very small
boundary layer of the wall, the two formulas 
\ref{p_wall_equ} and \ref{equ_osmosis_pressure}
give the same values for the osmotic pressure, up to a small difference
which %PRK: removed word
 vanishes as the width of the boundary layer is taken 
to zero.
For potentials which 
are "soft-walled" in the sense that solute molecules interact with 
the wall on a length scale comparable
to the magnitude of the 
fluctuations of the size of the solute molecules, as in 
Subsection \ref{section_application_pair_osmotic}, 
the average pressure of 
the fluid and the average pressure exerted on the wall may 
in fact differ.  In related work, 
 we are investigating the various 
 pressures associated with osmotic phenomena 
and exploring the influence of finite wall and molecule sizes
~\citep{atzberger2006d}.  For a discussion 
of how the osmotic pressure can be used
to drive fluid 
flow in a mesoscopic pump, see the related work~\citep{atzberger2005c}.
We now present a few examples to demonstrate how more 
complex structures immersed in the fluid 
can be simulated with the stochastic immersed boundary method
and to show how the osmotic pressure
associated with wall forces can be derived from the thermal 
fluctuations of these structures as simulated by the 
method. 

Before proceeding in the following subsections 
to consider osmotic pressure effects of more complex structures,
we remark that the
wall pressure can be computed from 
the average of the radial confinement 
force $ \mathbf{f}_{\mathrm{conf}} 
(\mathbf{x}) = - f_{\mathrm{conf}}(|\mathbf{x}|) \mathbf{x}/{|\mathbf{x}|}$ 
acting on the particles
and equation \ref{osmosis_p_wall_equ} in the 
case of a spherical chamber.  
From Newton's third law (principle of equal and opposite forces):
\begin{eqnarray}
\mathbf{f}_{\mathrm{conf}} (\mathbf{x}) =
 \int_{\partial \Omega}
-\mathbf{F}_{\mathrm{pw}}(\mathbf{y} - \mathbf{x}) \,
d\mathbf{y},
\end{eqnarray}
re-expressed using the isotropy of the forces involved, we have: %PRK
\begin{eqnarray}
-f_{\mathrm{conf}}(|\mathbf{x}|)\frac{\mathbf{x}}{|\mathbf{x}|}
& = & 
\int_{\partial \Omega} -q(|\mathbf{y} - \mathbf{x}|) 
\frac{\mathbf{y} - \mathbf{x}}{|\mathbf{y} - \mathbf{x}|}
d{\mathbf{y}}.
\end{eqnarray}
In spherical coordinates, this can be written:
\begin{eqnarray}
f_{\mathrm{conf}}(r) & = & \frac{\pi R}{r^2} \int_{R - r}^{R + r} 
q(\rho) (r^2 - R^2 + \rho^2) d\rho.
\end{eqnarray}

If the model for the solute-wall interaction force can be assumed to vanish at separation 
distances comparable to the chamber radius, then this integral relation between $ q $ and 
$ f $ can be inverted to obtain:
\begin{eqnarray}
\label{q_inversion_formula}
\\
\nonumber
q(\rho) & = & \left(
\frac{1}{2{\pi R}\rho^2}
\right)
\left(\rho (R-\rho) f^{\prime}_{\mathrm{conf}} (R-\rho) + (R+\rho) 
f_{\mathrm{conf}}(R-\rho) + \int_{0}^{R-\rho} f_{\mathrm{conf}}(s) ds
\right).
\end{eqnarray}  
The pressure on the wall can then be computed from the effective bulk confinement force 
$ f_{\mathrm{conf}} (r) $ using \ref{p_wall_equ}, \ref{def_h_r}, and this inversion
formula.

\subsection{Application: Simulation of Interacting Immersed Particles 
and Osmotic Pressure}
\label{section_application_pair_osmotic}

We now discuss application of the stochastic immersed boundary method
in determining the osmotic pressure when the confined particles can 
interact.  In particular, we consider the case in which particles 
interact in distinct pairs (dimers) through a spring with 
non-zero rest length and are confined to an approximately 
$400 \mbox{nm}$ spherical 
chamber.  Note 
that the solute 
particles are confined in a microscopic chamber, in the sense that 
the chamber diameter is comparable or smaller than the length-scale 
associated with the solute particle interactions between the monomers.
This is in contrast to a macroscopic chamber in which solute particles 
interact on a length-scale very much smaller than the chamber diameter 
and where the van't Hoff law is well established with the osmotic 
pressure depending only on the number of solute particles and not 
on their physical characteristics.  For example, in microscopic chambers 
the amplitude of the fluctuations of a solute particle's diameter may be 
comparable to the chamber diameter and play a non-negligible role in the 
osmotic pressure associated with confinement.  

To investigate these effects, 
we consider how the osmotic pressure 
changes as the binding strength for a collection of dimers is varied.  
From the classical van't Hoff's law, it would be expected that the osmotic 
pressure for tightly bound dimers are half that of the zero-binding case 
(free monomers), because a tightly bound dimer behaves effectively as a single 
particle.  This concept is exploited to suggest the design of an 
osmotically driven pumping apparatus in~\citep{atzberger2005c}.  From 
simulations using the stochastic immersed boundary method, we can 
investigate how the osmotic pressure varies between the unbound 
and tightly bound regimes.

\begin{figure}[h*]
\centering
\epsfxsize = 1.5in
\epsffile[14 14 225 237]{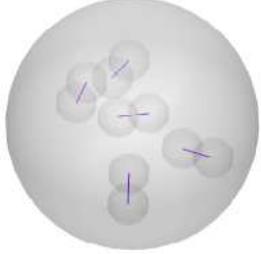}
\caption[Pair Confined]
{Illustration of five distinct pairs of coupled particles confined in a 
spherical chamber with the soft-wall potential given by 
equation \ref{equ_pairConfine}.
Simulations were performed for five pairs 
of particles coupled with interaction energy given by equation 
\ref{equ_pairInteraction}.}
\label{figure_pairConfined}
\end{figure}

\begin{figure}[h*]
\centering
\epsfxsize = 4.3in
\epsffile[14 14 375 233]{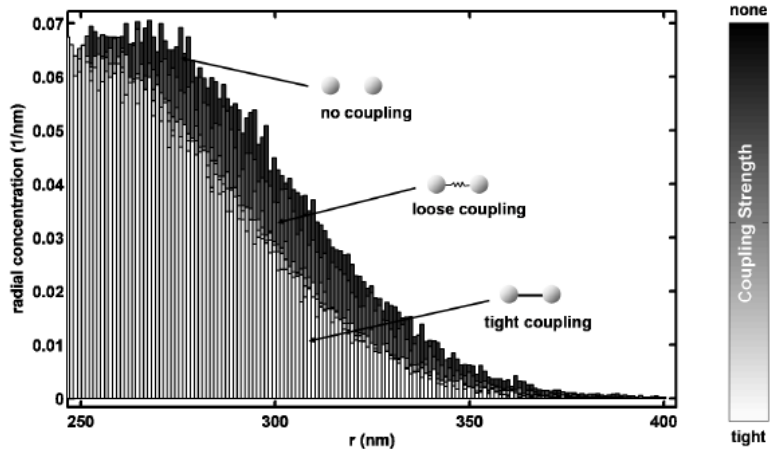}
\caption[Radial Force Density of Coupled Pairs]
{The geometrically weighted particle concentration 
$c(|\mathbf{x}| = r)r^2 $ 
of the monomers in the spherical shell of radius $r$ used in equation 
\ref{p_wall_equ}.  As the coupling is tightened, the particle concentration %PRK
has more weight at 
smaller radii.  The concentration is shown only over the boundary layer 
over which the confinement force is exerted.}
\label{figure_pairRadialForceDensity}
\end{figure}

Each pair of particles is coupled by a potential energy corresponding to 
a standard spring model with finite rest length $ \ell $:
\begin{eqnarray}
\label{equ_pairInteraction}
\Phi_{2}(\mathbf{X}_1, \mathbf{X}_2) = 
\frac{K}{2}\left(|\mathbf{X}_1 - \mathbf{X}_2| - \ell\right)^2, 
\end{eqnarray}
where $\mathbf{X}_1 $ and $ \mathbf{X}_2$ denote 
the particle locations and $K$ represents the 
spring stiffness.  Each monomer at location $\mathbf{x}$ 
is subject to a confinement force given by the radially %PRK
 symmetric potential:
\begin{eqnarray}
\label{equ_pairConfine}
\Phi_{1}(\mathbf{x}) = 
\left\{
\begin{array}{ll}
0, & |\mathbf{x}| \leq R_{1}, \\
\frac{C_0}{2}\left(|\mathbf{x}| - R_1\right)^2, & R_{1} \leq |\mathbf{x}| \leq R_2 \\
\frac{C_0}{2}\left(R_{2} - R_1\right)^2, &|\mathbf{x}| \geq R_{2}.
\end{array}
\right.
\end{eqnarray}
%PRK:  modified equation
This potential can be thought of as arising from the interaction force 
of the confined solute monomers with particles distributed uniformly 
over the walls of a spherical chamber having radius $R=R_{2}$.  %PRK
The 
formula given in equation \ref{q_inversion_formula}
gives the relationship between the monomer-wall 
interaction force and the radial confinement force.
For the simulations the parameters were chosen as 
$\ell = 100 \mbox{nm}$, $R_1 = 375 \mbox{nm}$, 
$R_2 = 400 \mbox{nm}$, 
$C_0 = 8k_B{T}/(R_2 - R_1)^2$.
The setup is depicted pictorially 
in Figure~\ref{figure_pairConfined}, and movies
of the simulations can be found in the Supplemental 
Materials.

\begin{figure}[h*]
\centering
\epsfxsize = 4in
\epsffile[0 0 362 286]{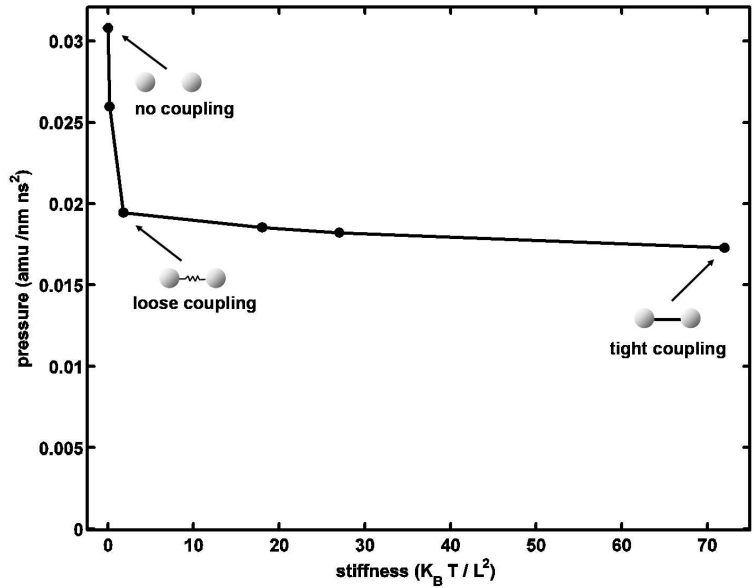}
\caption[Osmotic Pressure of Coupled Pairs]
{The osmotic pressure for confined dimers as a function of coupling strength.
As the coupling strength of the dimers increases, the
osmotic pressure decreases in a non-linear manner.   The 
osmotic pressure value for unbound monomers is approximately
double that of the the tightly bound monomers, which 
is in accordance with what is expected under van't Hoff's law.
Deviations from van't  Hoff's law are apparent for intermediate coupling strengths 
for which the length scale of the dimers is comparable to those of the chamber and the wall thickness.  
}
\label{figure_pairOsmoticPressure}
\end{figure}

Figure \ref{figure_pairRadialForceDensity} displays 
the results of simulations with the stochastic immersed boundary method that  
show that as the stiffness is increased, the particle density decreases
 for each radius $r$ within the region of the confining potential.  
   The pressure consequently 
drops with increasing coupling stiffness, as shown in 
Figure \ref{figure_pairOsmoticPressure}.  We see also that,  in accordance 
with van't Hoff's law,  the pressure in the strong coupling limit, where the 
particle pairs behave effectively as single entities, is cut to roughtly half from 
the no coupling case.  We observe deviations from the classical van't Hoff law 
when the length scale of the bound molecules is comparable to that of the wall
or the chamber. 

An intuitive statistical explanation for the pressure 
drop is that a strongly coupled particle is less likely to venture far into 
the confining potential because roughly half the time its partner will 
encounter the confining potential first, be repelled, and pull its 
accompanying particle back away from the confining potential sooner 
than it would have on its own. 
In more physical terms,  the entropy of the particle pairs 
decreases and consequently the entropic penalty associated with 
confinement is reduced as the coupling strength increases.
For  coupling values that make the dimer length scale comparable to the 
microscopic chamber size, the osmotic pressure assumes an intermediate 
value which  is not well described by a van't Hoff's law.

\subsection{Application: Simulation of Polymer Chains and Polymer Knots}
\label{section_application_knots}

A fundamental feature of the stochastic immersed boundary method is that 
each structure evolves according to a local average of a common fluid 
velocity field.
The method therefore automatically captures the physical phenomenon that
the velocities of immersed structures become strongly correlated when they
are close together 
in space. 
Mathematically, 
the solution map of the immersed 
structures and surrounding fluid volume, which maps a 
configuration of the fluid and structures at a reference time 
to the solution configuration at a later 
time $t$, can be viewed as a homeomorphism.  
Consequently, in the continuous-time framework, 
the method preserves topological invariants of the immersed 
structures, such as the knottedness of a continuous closed 
curve, as they evolve.  This is
in contrast to other simulation methods, such as Stokesian 
Dynamics~\citep{jfb:sd,as:asds,ts:mms}, which would require 
explicit excluded volume constraints and/or repulsion forces
between monomers to prevent topological changes.

\begin{figure}[h*]
\centering
\epsfxsize = 4in
\epsffile[0 0 372 312]{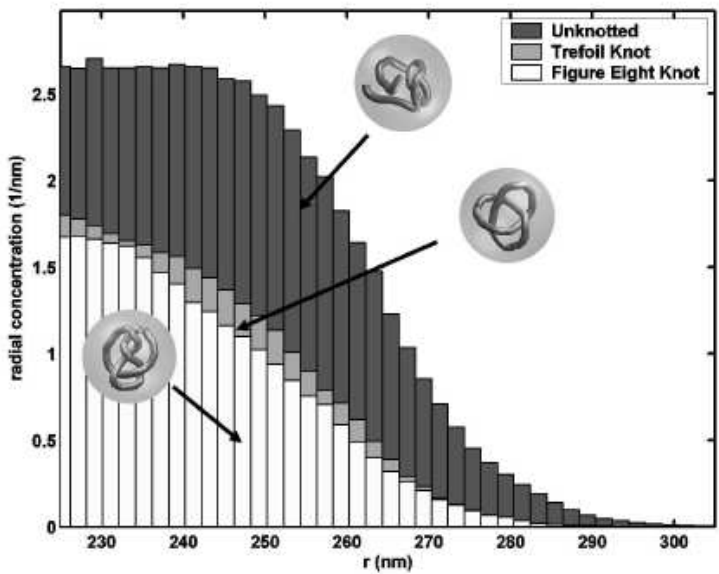}
\caption[Osmotic Pressure of Knots]
{The bar graph shows the average concentration of 
monomers within the chamber in the boundary 
layer in which they interact with the confinement forces 
(Subsection \ref{section_application_osmotic}). 
As the knottedness of the polymer increases, 
the monomers are more restricted and concentrate on average 
toward the chamber center spending less time interacting 
with the confinement forces.  As a consequence the 
concentration of monomers in the boundary layer decreases
and the overall pressure drops (Table \ref{table_knotOsmoticPressure}). 
The parameters in the simulations were taken the same as 
in Subsection \ref{section_osmoticNoninteracting}.}
\label{figure_knotRadialForceDensity}
\end{figure}

To demonstrate this feature of the method in practice (with finite time step)
and to show how worm-like chain polymers can be simulated, 
the stochastic immersed boundary method was applied
for a generic polymer chain, a polymer
trefoil knot, and a polymer figure eight knot.  From these simulations
the osmotic pressure of confinement was estimated for each
of the polymers.  
The results of the average concentration of the 
polymer monomers, which 
determine the average radial force density 
exerted on the confining wall,
is given in Figure \ref{figure_knotRadialForceDensity}.

As the knottedness 
of the polymer increases, its constituent monomers spend less time at 
large radii, and as seen in Table \ref{table_knotOsmoticPressure}, the  
osmotic pressure is significantly reduced.  An intuitive 
explanation is that as the knottedness of the polymer increases,
this restricts the intrinsic configurations 
accessible to the thermally fluctuating polymer.  In physical
terms, the knottedness reduces the entropic penalty of confining 
of the polymer.  Movies showing simulations of the thermally fluctuating
polymer knots can be found in the Supplemental Materials.

\begin{table}[h*]
\centering
\caption{Osmotic Pressure of Polymer Knots}
\vspace{0.5cm}
\begin{tabular}{|l|l|}
\hline
Knot Type & Osmotic Pressure ($\mbox{amu}/\mbox{nm}\cdot\mbox{ns}^2$)  \\
\hline
Unknotted & 0.16 \\
Trefoil Knot & 0.0439 \\
Figure Eight Knot & 0.0392 \\
\hline
\end{tabular}
\label{table_knotOsmoticPressure}
\end{table}

\subsection{Application: Simulation of a Basic Model for a 
Molecular Motor Protein
Transporting a Membrane-Bound 
Cargo Vesicle  
}
\label{section_application_motor}

\begin{figure}[h*]
\centering
\epsfxsize = 5.5in
\epsffile[14 14 375 126]{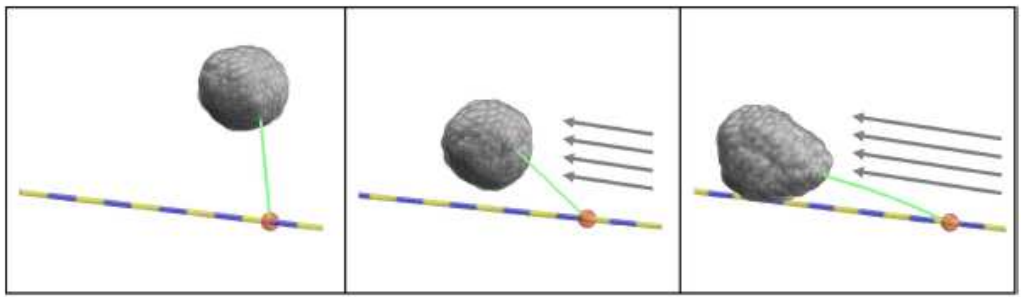}
\caption[Ratchet Model Subject to Hydrodynamic Drag]
{Illustration of the basic model of a molecular motor protein 
immersed in a fluid and towing a cargo vesicle.  As the fluid flow
strengthens, the hydrodynamic drag on the cargo increases
and a load force is exerted in opposition to the 
motor transport.  For large opposing fluid flows, the 
cargo may significantly change shape in response to the 
flow. %PJA: rephrased
}
\label{figure_motorModel}
\end{figure}

We now discuss how more complex systems can be simulated with 
the immersed boundary method.  On a subcellular level 
motor proteins interact with cytoskeletal structures, 
such as actin and microtubules, to generate force and to 
transport materials within the cell.  For example, 
neurotransmitters are produced in the cell body of neurons
and transported by kinesin motor proteins 
along axons to the vicinity of the synaptic cleft where they are
packaged for future release 
~\citep{thecell}.  We demonstrate how 
the stochastic immersed boundary method can be applied to 
simulate a basic model of a molecular motor protein 
immersed in a fluid
moving along a filament which transports a cargo vesicle
(Figure \ref{figure_motorModel}).

\begin{figure}[h*]
\centering
\epsfxsize = 4in
\epsffile[14 14 375 278]{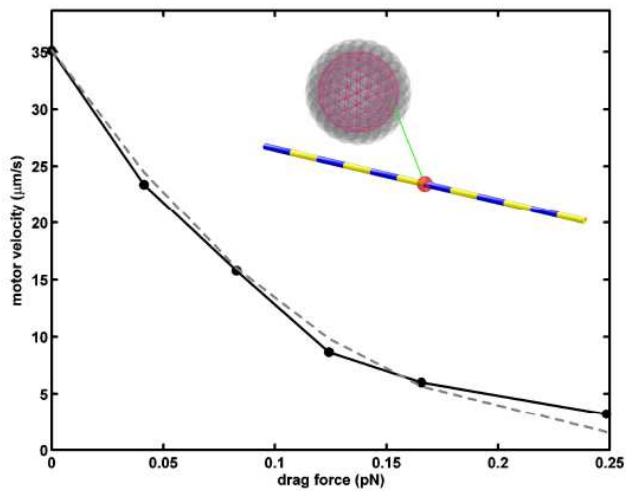}
\caption[Ratchet Force Velocity]
{The mean motor velocity vs the hydrodynamic drag force.  
The data points joined by the solid curve shows the mean motor velocity 
${X_{\tau}}/{\tau}$ obtained from simulations over approximately $\tau = 3 \mbox{ms}$.
The dashed curve shows the mean velocity for an idealized ratchet
having a viscous drag comparable to the spherical vesicle and subject 
to a constant load force the same strength as the hydrodynamic drag 
force~\cite{peskin1993, jh:mmpc}.}
\label{figure_motorForceVelocity}
\end{figure}

The motor protein is modeled as a Brownian 
Ratchet~\cite{peskin1993, jh:mmpc} and the cargo vesicle is modeled by a 
triangulated mesh which forms a membrane enclosing a spherical 
volume.  The nodes of the mesh are linked together by 
springs of the form given in equation \ref{equ_pairInteraction} with
non-zero rest lengths determined by the distance
between nodes in an initial spherical configuration.
The cargo is linked at the vesicle surface to the motor by a 
spring of the form given by equation \ref{equ_pairInteraction} 
with a non-zero rest length of approximately $100 \mbox{ nm}$.
The spherical vesicle has a radius of $125 \mbox{nm}$ and the
ratcheting intervals (light and dark inset of 
Figure \ref{figure_motorForceVelocity}) are of 
length $100 \mbox{ nm}$. 

To examine how the mean velocity of transport by the 
motor protein behaves under different loading conditions, 
simulations were performed in which a hydrodynamic 
drag force is generated on the vesicle cargo by a 
bulk flow of the fluid~\citep{jh:mmpc,thecell}.  
The mean velocity for 
different strengths of the countering fluid flow is
plotted in Figure~\ref{figure_motorForceVelocity},
where we see that for significantly large 
opposing flow, the motor can almost 
be made to ``stall''.  
The stochastic immersed boundary method allows for hydrodynamic 
effects associated with the shape and deformation of cargos to 
be investigated.  These effects, not typically considered in 
other numerical simulation approaches for molecular motors, 
may have important consequences for motor/polymerization ratchet 
transport when 
more details of the motor are taken into account 
or with 
 more complex cargos 
  such as membrane 
tubes, small cell organelles, or 
chromosomes~\citep{thecell,raj2006,koster2003}.  
More sophisticated models can also be formulated
within the stochastic immersed boundary method 
framework, such as the case in which multiple
motor proteins transport a common cargo, interact 
by crosslink cytoskeletal filaments, or include additional
mechanical degrees of freedom of the motor protein itself.  
Some movies of our motor protein model transport simulations 
can be found in the Supplemental Materials.

\section{Conclusion and Discussion}
In this work we have discussed how thermal fluctuations
can be incorporated into the immersed boundary method
in a manner consistent with the laws of statistical mechanics.  
A new stochastic numerical method was proposed that allows for
a range of time steps to be taken in which the 
fastest degrees of freedom of the fluid-particle system 
are either underresolved, partially resolved, or fully resolved.
In addition, the numerical method was designed to take 
into account in a systematic way the statistical contributions 
of the thermal fluctuations over long time steps with the 
correct correlations between the particles and fluid.  

To investigate the behavior of the immersed boundary
framework and the stochastic numerical method 
with respect to well-known laws in statistical physics
a number of theoretical results were obtained for the method and 
compared with numerical simulations.  In particular, it was
shown that immersed particles simulated 
with the numerical method 
exhibit the correct scaling in the physical parameters 
for the mean squared displacement in three dimensions.  
It was also shown that the stochastic
numerical method captured 
inertial effects of the fluid with a velocity autocorrelation function
for a particle that for long times decays 
with algebraic order $\tau^{-3/2}$.  We
further found that particles appear to have the correct Boltzmann equilibrium 
statistics.  Moreover, the method was found to produce 
the van't Hoff law of osmosis for a particle confined to a spherical
chamber recovering the correct osmotic pressure.  In addition results
were presented which showed how the osmotic pressure could be computed
for interacting pairs of particles and worm-like chain polymers, including
a trefoil and figure eight polymer knot.  A more complex application
 of a basic model of a molecular motor protein immersed in a 
fluid and towing a vesicle-bound cargo subject to a hydrodynamic drag 
force was simulated and the force-velocity statistics computed.

These basic physical checks indicate that the stochastic immersed boundary 
method has the capability to capture many important features of 
thermally fluctuating systems involving immersed structures which 
interact with a fluid.  The results presented suggest the promise of 
the stochastic immersed boundary method as an effective approach in 
modeling and simulating the mechanics of biological systems at the 
cellular and intracellular level.

\section*{Acknowledgments}
The author P.J.A was supported by NSF VIGRE Postdoctoral Research
Fellowship Grant DMS - 9983646
and NSF Mathematical Biology Grant DMS - 0635535.  The authors would
like to thank Eric Vanden-Eijnden and David Cai for helpful 
discussions on analytical and physical aspects of the work,
and Tom Bringley for a careful reading of a 
preliminary draft.  The authors also thank David McQueen for 
discussions concerning the immersed boundary method and Yuri Lov 
for lending computational resources used to obtain some of 
the numerical results.  We are especially indebted to George Oster,
whose vision of direct numerical simulation of osmotic phenomena
via random forces inspired this work.

\begin{appendix}

\section{The Representation Function $\delta_a$ for Immersed Particles}
\label{appendix_delta_func}
In the immersed boundary method,
 it is required
that a function $\delta_a$ be specified to represent the 
elementary 
particles.  The representation of this 
function is 
often derived from the following function $\phi$ 
which is known to have desirable numerical properties 
~\citep{peskin2002}: 
\begin{eqnarray}
\label{equ_phi_def}
\phi(r) & = & \left\{
\begin{array}{ll}
0
                  & \mbox{, if $r \leq -2$} \\
                  & \\
\frac{1}{8} \left(5 + 2r - \sqrt{-7 - 12r - 4r^2} \right) 
                  & \mbox{, if $-2 \leq r \leq -1$} \\
                  & \\
\frac{1}{8} \left(3 + 2r + \sqrt{1 - 4r - 4r^2} \right) 
                  & \mbox{, if $-1 \leq r \leq 0$} \\
                  & \\
\frac{1}{8} \left(3 - 2r + \sqrt{1 + 4r - 4r^2} \right) 
                  & \mbox{, if $0 \leq r \leq 1$} \\
                  & \\
\frac{1}{8} \left(5 - 2r - \sqrt{-7 + 12r - 4r^2} \right) 
                  & \mbox{, if $1 \leq r \leq 2$} \\
                  & \\
0
                  & \mbox{, if $2 \leq r$.} 
                  \\
\end{array}
\right.
\end{eqnarray}

For three dimensional systems the function 
$\delta_a$ representing elementary 
 particles of size $a$ is  
\begin{eqnarray}
\label{equ_phi_def_delta_a}
\delta_a(\mathbf{r}) = \frac{1}{a^3}
\phi\left(\frac{\mathbf{r}^{(1)}}{a}\right)
\phi\left(\frac{\mathbf{r}^{(2)}}{a}\right)
\phi\left(\frac{\mathbf{r}^{(3)}}{a}\right), 
\end{eqnarray}
where the superscript indicates the index of the vector component.

To maintain good numerical properties, 
 the particles are 
restricted to sizes $a = n\Delta{x}$, where $n$ is a positive 
integer.  For a derivation and a detailed discussion of 
the properties of these functions 
see ~\citep{peskin2002}.

\section{The Fourier Coefficients of the
Function $\delta_a$ Used to Represent an Immersed Particle}
\label{appendix_delta_func_hat}
Throughout the paper it will be useful to consider the
Fourier coefficients of the
function $\delta_a(\mathbf{x} - \mathbf{X})$
used to represent an elementary 
 particle situated 
at position $ \mathbf{X}$.  While the
function is defined for all $\mathbf{x} \in \Lambda$,
it is often useful to consider
the restriction of the function to the discrete
lattice points $\{\mathbf{x}_{\mathbf{m}} =
\mathbf{m}\Delta{x} | \mathbf{m} \in \mathbb{Z}_N^3\}$.

We will use the following notation to denote the discrete 
Fourier transform of the
delta function restricted to the lattice:
\begin{eqnarray}
\hat{\delta}_{a,\mathbf{k}}(\mathbf{X})
& = & \frac{1}{N^3} \sum_{\mathbf{m}} \delta_{a}
(\mathbf{x}_{\mathbf{m}} - \mathbf{X})
\exp\left(-i2\pi{\mathbf{k}}\cdot\mathbf{m}/N\right).
\end{eqnarray}
The dependence of the Fourier coefficients on the
particle position $ \mathbf{X} $ (relative to the lattice)
is explicitly noted.
When the dependence on $ \mathbf{X} $ is not explicitly noted,
then we will be referring implicitly 
to the discrete Fourier 
transform of the delta function when centered on a 
lattice point:  $\hat{\delta}_{a,\mathbf{k}} :=
\hat{\delta}_{a,\mathbf{k}}(\mathbf{0})$.

\section{Autocorrelation Function for the Velocity Field of the Fluid}
\label{appendix_velocity_autocorrelation}
In this section the autocorrelation function is computed 
for the velocity field of the fluid in the absence of force
$\mathbf{f}_{\mbox{\small prt}} = 0$.  
This is done 
by representing the velocity field in
Fourier space and computing the autocorrelation function 
of each mode $\mathbf{k}$.  From 
equation \ref{analytic_sol_u} and
standard stochastic calculus 
the steady-state 
autocorrelation function 
of the $\mathbf{k}^{th}$ mode
when $s > r$ is
\begin{eqnarray}
\label{equ_corr_deriv_1_a}
E\left(\overline{\hat{\mathbf{u}}_{\mathbf{k}}(s)}
\cdot\hat{\mathbf{u}}_{\mathbf{k}}(r)\right) &&
\\
\label{equ_corr_deriv_1_b}
= 2D_{\mathbf{k}} E\left(\overline{\int_{-\infty}^{s} 
e^{-\alpha_{\mathbf{k}}(r - w)}
\wp_{\mathbf{k}}^{\perp}d\mathbf{\tilde{B}}_{\mathbf{k}}(w)}
\right.
&&
\left.
\cdot
 \int_{-\infty}^{r}  
e^{-\alpha_{\mathbf{k}}(s - q)}
\wp_{\mathbf{k}}^{\perp}{d\mathbf{\tilde{B}}_{\mathbf{k}}(q)}
\right), 
% = 
%\label{equ_corr_deriv_1_c}
%2D_{\mathbf{k}} E\left( 
%\overline{\int_{-\infty}^{r}  
%e^{-\alpha_{\mathbf{k}}(r - w)}
%\wp_{\mathbf{k}}^{\perp}
%d\mathbf{\tilde{B}}_{\mathbf{k}}(w) }
%\cdot
%\right.
%&&
%\left.
%\int_{-\infty}^{r}  
%e^{-\alpha_{\mathbf{k}}(s - q)}
%\wp_{\mathbf{k}}^{\perp}
%d\mathbf{\tilde{B}}_{\mathbf{k}}(q)
%\right) \\
%\nonumber
\end{eqnarray}
where the notation $\wp_{\mathbf{k}}^{\perp}$ 
denotes projection orthogonal to $\hat{\mathbf{g}}_{\mathbf{k}}$ as
defined in Subsection \ref{section_fluid_fourier}.
%The last equality follows from
%\begin{eqnarray}
%E\left( 
%\overline{
%\int_{r}^{s}  
%e^{-\alpha_{\mathbf{k}}(r - w)}
%\wp_{\mathbf{k}}^{\perp}d\mathbf{\tilde{B}}_{\mathbf{k}}(\mathbf{w})}
%\cdot
%\int_{-\infty}^{r}  
%e^{-\alpha_{\mathbf{k}}(s - q)}
%\wp_{\mathbf{k}}^{\perp}d\mathbf{\tilde{B}}_{\mathbf{k}}(\mathbf{q})
%\right) & = & 0 \\
%\end{eqnarray}
%where it is used that Ito integrals when 
%integrating deterministic functions over different ranges in time
%are independent.

By applying Ito's Isometry to 
\ref{equ_corr_deriv_1_b}, and observing the symmetry under the 
interchange $ s \leftrightarrow r $, the autocorrelation 
function is given by
\begin{eqnarray}
\label{equ_corr_deriv_2}
E\left(\overline{\hat{\mathbf{u}}_{\mathbf{k}}(s)}
\cdot\hat{\mathbf{u}}_{\mathbf{k}}(r)\right)
& = & 
\left\{
\begin{array}{ll}
3\frac{D_{\mathbf{k}}}{\alpha_{\mathbf{k}}}
%2D_{\mathbf{k}}2[2]_{\mathbf{k}}
e^{-\alpha_{\mathbf{k}}|s - r|}
%\int_{-\infty}^{r}  
%e^{-2a_{\mathbf{k}}(r - q)}
%dq 
& \mbox{if $\mathbf{k} \in \mathcal{K}$} \\
4\frac{D_{\mathbf{k}}}{\alpha_{\mathbf{k}}}
%2D_{\mathbf{k}}2[2]_{\mathbf{k}}
e^{-\alpha_{\mathbf{k}}|s - r|}
%\int_{-\infty}^{r}  
%e^{-2a_{\mathbf{k}}(r - q)}
%dq 
& \mbox{if $\mathbf{k} \not\in \mathcal{K}$} \\
\end{array}
\right. \\
%\nonumber
%& = & 
%\Upsilon_{\mathbf{k}}
%%2 [2]_{\mathbf{k}}
%\frac{D_{\mathbf{k}}}{a_{\mathbf{k}}}
%e^{-a_{\mathbf{k}}|s - r|} \\
\nonumber
& = & 
\Upsilon_{\mathbf{k}}
\frac{k_B{T}}{\rho L^3}
e^{-\alpha_{\mathbf{k}}|s - r|}, 
\end{eqnarray}
where
\begin{eqnarray}
\Upsilon_{\mathbf{k}} = 
\left\{
\begin{array}{ll}
3,
& \mbox{ $\mathbf{k} \in \mathcal{K}$} \\
2,
& \mbox{ $\mathbf{k} \not\in \mathcal{K}$}, 
 \\
\end{array}
\right.
\end{eqnarray}
and the index set $\mathcal{K}$ is defined in \ref{equ_def_K}.

The factor $\Upsilon_{\mathbf{k}}$ arises from the incompressibility
constraint \ref{equ_DFT_incompressible}, the real-valuedness 
constraint \ref{equ_DFT_real_constr},
and the dimensionality of 
the space orthogonal to $\hat{\mathbf{g}}_{\mathbf{k}}$.
See Subsection \ref{section_thermal_forcing} for a discussion
of how the constraints affect 
 $D_\mathbf{k}$.

The spatio-temporal correlation function of the velocity field $\mathbf{u} $
is then given by 
\begin{eqnarray}
\nonumber
E\left(\mathbf{u}_{\mathbf{m}}(s)\cdot
\mathbf{u}_{\mathbf{n}}(r)\right) 
& = & \sum_{\mathbf{k}} \sum_{\mathbf{k'}} 
E\left(\overline{\hat{\mathbf{u}}_{\mathbf{k}}(s)}\cdot
\mathbf{u}_{\mathbf{k'}}(r)\right)
\exp\left({i2\pi(\mathbf{n}\cdot \mathbf{k'} 
- \mathbf{m}\cdot \mathbf{k})/N}\right) \\
\label{equ_autocorrolation_1}
& = & \sum_{\mathbf{k}} 
E\left(\overline{\hat{\mathbf{u}}_{\mathbf{k}}(s)}\cdot
\hat{\mathbf{u}}_{\mathbf{k}}(r)\right)
\exp\left({i2\pi(\mathbf{n}- \mathbf{m}) \cdot \mathbf{k}/N}\right) \\
\nonumber
& = & \frac{k_B{T}}{\rho L^3}
\sum_{\mathbf{k}} \Upsilon_{\mathbf{k}}
e^{-a_{\mathbf{k}}|s - r|}
\exp\left({i2\pi(\mathbf{n}- \mathbf{m}) \cdot \mathbf{k}/N}\right).
\end{eqnarray}
To obtain the second equality, we used the statistical 
independence of the Fourier modes of the velocity field 
when the indices $ \mathbf{k} $ and $ \mathbf{k}^{\prime} $ 
are distinct and do not correspond to conjugate modes 
(see \ref{equ_DFT_real_constr}).   When the indices 
$ \mathbf{k} $ and $ \mathbf{k}^{\prime} $ do refer 
to conjugate but distinct modes, then the average vanishes 
because a mean zero random variable $ Z $ with independent 
and identically distributed real and imaginary components 
satisfies $ \langle Z^{2} \rangle = 0$.     The last equality 
follows by substitution from equation 
\ref{equ_corr_deriv_2}.

\section{Constants: Accuracy and Error Estimates}
\label{appendix_constants_accuracy_error}
The non-dimensional factors $Q$ appearing in the error estimates in
Section~\ref{section_accuracy} are approximately independent of the
physical parameters.
For comparison of the theoretical error 
estimates with numerical simulations, it is useful to 
compute the factors for specific physical parameters to 
obtain estimated values.  Evaluation of the expressions for the
$ Q $ constants for  the system with 
parameters given in Table \ref{table_params_diffusion_compare}
gives the following values:
\begin{eqnarray}
Q_1 & = & 0.563 \\
Q_2 & = & 7.87. 
\end{eqnarray}

% Q_1 = {\rho a^4}/{k_B{T}} C_1
% Q_2 = {a/D} C_2

\clearpage
\pagebreak
\newpage

\end{appendix}

\clearpage
\pagebreak
\newpage

%\bibliography{draft7}
\ifx\undefined\allcaps\def\allcaps#1{#1}\fi\newcommand{\noopsort}[1]{}
  \newcommand{\printfirst}[2]{#1} \newcommand{\singleletter}[1]{#1}
  \newcommand{\switchargs}[2]{#2#1}
  \ifx\undefined\allcaps\def\allcaps#1{#1}\fi\def\cprime{$'$}

\clearpage
\pagebreak
\newpage

\section*{Tables}

\clearpage
\pagebreak
\newpage

\begin{table}[h]
\centering
\caption{Parameters of the Method}
\vspace{0.5cm}
\label{table_3Dsim_param_descr}
\begin{tabular}{|l|l|}
\hline
Parameter & Description \\
\hline
$k_B$              & Boltzmann's constant \\
$T$                & Temperature \\
$L$                & Period Length of Fluid Domain \\
$\mu$              & Fluid Dynamic Viscosity \\
$\rho$             & Fluid Density \\
$N$                & Number of Grid Points in each Dimension \\
$\Delta{t}$        & Time Step \\
$\Delta{x}$        & Space Between Grid Points \\
$a$                & Effective Elementary Particle Size (approximate radius)\\
\hline
\end{tabular}
\end{table}

\begin{table}[h]
\centering
\caption{Values used in Numerical Simulations}
\vspace{0.5cm}
\label{table_params_diffusion_compare}
\begin{tabular}{|l|l|}
\hline
Parameter & Description \\
\hline
$T$                & $300                \mbox{ K}$ \\
$L$                & $1000               \mbox{ nm}$ \\
$\mu$              & $6.02 \times 10^5 \mbox{ amu/(nm $\cdot$ ns)}$\\
$\rho$             & $602 \mbox{ amu/$\mbox{nm}^3$}$ \\
$N$                & 32 \\
%$\Delta{t}$        & 1000 \mbox{ns} \\
\hline
\end{tabular}
\end{table}

\clearpage
\pagebreak
\newpage

\clearpage
\pagebreak
\newpage

\begin{table}[h]
\centering
\caption{Notation Conventions}
\vspace{0.5cm}
\begin{tabular}{|l|l|}
\hline
Parameter & Description \\
\hline
$\delta_a$                         & Representation function of an immersed elementary particle
                                     of size $a$ \\ 
$\delta_{a,\mathbf{k}}$            & The $\mathbf{k}^{th}$ Fourier coefficient of the 
                                     particle representation function \\
$\alpha_{\mathbf{k}}$              & Damping of the $\mathbf{k}^{th}$ Fourier mode \\
$D_{\mathbf{k}}$                   & Strength of the thermal forcing of the $\mathbf{k}^{th}$ 
                                     Fourier mode \\
$\mathbf{u}_m$                     & Fluid velocity at the $\mathbf{m}^{th}$ grid point \\
$\hat{\mathbf{u}}_k$               & The $\mathbf{k}^{th}$ Fourier mode of the fluid 
                                     velocity field \\
$\mathbf{U}$                       & Smoothed fluid velocity field for 
                                     immersed elementary 
                                     particles \\
$\mathbf{x}_m$                     & Position vector of the $\mathbf{m}^{th}$ Eulerian grid 
                                     point \\
$\mathbf{X}^{[j]}$                 & Position vector of the $j^{th}$ 
                                     immersed elementary particle \\
$\mathbf{f}_{\small \mbox{\small prt}}$ & Force 
                                          density arising from the immersed structures \\
$\mathbf{f}_{\small \mbox{\small thm}}$    & Force density arising from the thermal forcing \\
$\hat{\mathbf{f}}_k$               & The $\mathbf{k}^{th}$ Fourier mode of the structural force density 
 field \\
\hline
\end{tabular}
\end{table}

\clearpage
\pagebreak
\newpage

\end{document}